\documentclass[12pt]{article}

\usepackage{latexsym}
\usepackage{amssymb}
\usepackage{amsmath}

\newtheorem{theorem}{Theorem}

\def\proof{\medbreak\noindent{\bf Proof}}

\def\{{\lbrace}
\def\}{\rbrace}

\def\cl{{\cal C}\!\ell}

\def\Com{{\rm Com}}

\def\com{{\rm Com}}
\def\ss{\stackrel}
\def\for{\quad\hbox{for}\quad}

\def\ww{\wedge\ldots\wedge}

\def\R{{\Bbb R}}
\def\C{{\Bbb C}}
\def\F{{\Bbb F}}

\def\U{{\rm U}}

\def\E{{\cal E}}
\def\Tr{{\rm Tr}}
\def\Pin{{\rm Pin}}
\def\diag{{\rm diag}}
\def\even{{\rm even}}
\def\odd{{\rm odd}}
\def\det{{\rm det}}

\def\Spin{{\rm Spin}}
\def\Pin{{\rm Pin}}

\def\be{\begin{equation}}
\def\ee{\end{equation}}

\def\Mat{{\rm Mat}}
\def\Pin{{\rm Pin}}

\newcommand{\dimension}{{\rm dim\,}}

\newcommand{\fin}{\par\hbox{$\blacksquare$}\medskip\par}

\newtheorem{lem}{Lemma.}

\begin{document}





\title{Unitary spaces on Clifford algebras}

\author{Marchuk N.G.\footnote{Steklov Mathematical Institute,
nmarchuk@mi.ras.ru}, Shirokov D.S.\footnote{Moscow State
Univercity, shirokov-dm@nm.ru}}

\begin{abstract}
For the complex Clifford algebra Cl(p,q) of dimension n=p+q we
define a Hermitian scalar product. This scalar product depends on
the signature (p,q) of Clifford algebra. So, we arrive at unitary
spaces on Clifford algebras. With the aid of Hermitian idempotents
we suggest a new construction of, so called, normal matrix
representations of Clifford algebra elements. These
representations take into account the structure of unitary space
on Clifford algebra.
\end{abstract}

\maketitle

\tableofcontents

\bigskip

Clifford algebras were invented by W.~K.~Clifford \cite{Clifford} in
1878 году. One of authors uses the Clifford algebra $\cl(1,3)$ in
the field theory \cite{Marchuk:AACA}-\cite{25.} (Dirac-Yang-Mills
equations). In the mentioned papers a notion of unitary space on
Clifford algebra was developed for the Clifford algebra $\cl(1,3)$.
Several structure equalities were found for $\cl(1,3)$.

In the present paper the notion of unitary space on Clifford algebra
is generalized for Clifford algebras $\cl(p,q)$ of dimensions
$n=p+q>4$ and for different signatures $(p,q)$. Also we prove
several structure equalities for the Clifford algebra (theorems
1-8).

With the aid of Hermitian idempotents (sections 6,7) we suggest a
new construction of, so called, normal matrix representations of
Clifford algebra elements. These representations take into account
the structure of unitary space on Clifford algebra.

Note that in papers
\cite{Hestenes, Lounesto,Snygg} there are several partial cases of the considered
structure equalities of the Clifford algebra. Namely, in \cite{Hestenes}
we see the following formula for the Clifford product of two Clifford algebra elements
$\ss{k}{U},\ss{l}{V}\in\cl(p,q)$, $p+q=n$ of ranks $k,l$
$$
\ss{k}{U}\ss{l}{V}=\ss{k-l}{W}+\ss{k-l+2}{W}+\ldots+\ss{k+l}{W},
$$
where $\ss{m}{W}=0$ for $m>n$ and for $m<0$.
This formula can be derived from Theorem 2 of the present paper.

Also, well known formulas
$$
e^a e_a=n,\quad e^a\ell e_a=(-1)^{n+1}n\ell,
$$
where $e^a$ are generators of Clifford algebra $\cl(p,q)$, $p+q=n$
and $\ell=e^1\ldots e^n$, can be considered as partial cases of
the proposition of Theorem 5.

Theorem 7 defines an operation of Hermitian conjugation $U\to
U^\dagger$ of Clifford algebra elements. In particular, for
$\cl(n)$ we get the operation $U^\dagger=U^*$ and for $\cl(1,n-1)$
we get the operation $U^\dagger=e^1 U^* e^1$. In both cases these
operations are well known in the literature (see, for example,
\cite{Snygg}).

\section{Clifford algebras}

Let $\F$ be the field of real numbers $\R$ or the field of complex
numbers $\C$ and let $n$ be a natural number. Consider the $2^n$
dimensional vector space $\E$ over the field $\F$ with a basis
\begin{equation}
e,\,e^a,\,e^{a_1 a_2},\ldots, e^{1\ldots n}, \quad
a_1<a_2<\cdots,
\label{basis:cl}
\end{equation}
with elements numbered by ordered multi-indices of length from $0$
to $n$. Indices $a,a_1,\ldots$ take values form $1$ to $n$. Let
$p,q$ be nonnegative integer numbers and $p+q=n$. Consider a
diagonal matrix of dimension $n$
\begin{equation}
\eta=\eta(p,q)=\diag(1,\ldots,1,-1,\ldots,-1)
\label{eta}
\end{equation}
with
$p$ pieces of $1$ and  $q$ pieces of $-1$ on the diagonal.
By $\eta^{ab}=\eta_{ab}$ denote elements of $\eta$. Following rules
define product
$U,V\to UV$ of elements of the vector space $\E$:
\begin{enumerate}
\item $\forall U,V,W\in\E$
\begin{eqnarray*}
&&U(V+W)=U V+U W,\quad
(U+V) W=U W+V W,\\
&&(U V) W=U(V W).
\end{eqnarray*}
\item $\forall U\in\E\,$ $e U=U e=U$.
\item $e^a  e^b+e^b e^a=2\eta^{ab}e$ for $a,b=1,\ldots,n$.
\item $e^{a_1}\ldots e^{a_k}=e^{a_1\ldots a_k}$ for
$1\leq a_1<\cdots<a_k\leq n$.
\end{enumerate}

This algebra is called
{\it the Clifford algebra} and denoted by
$\cl^\F(p,q)$ (if $\F=\C$, then
$\cl(p,q)=\cl^\C(p,q)$). For $p=n, q=0$ we use notation
$\cl^\F(n)=\cl^\F(n,0)$. Elements $e^a$ is called {\it generators} of
Clifford algebra
$\cl^\F(p,q)$.
$\cl^\R(p,q)$ is real Clifford algebra and $\cl(p,q)$ is complex
Clifford algebra.

Any element
$U\in\cl^\F(p,q)$ can be written in the form
\begin{equation}
U=ue+u_a e^a+\sum_{a_1<a_2} u_{a_1 a_2}e^{a_1 a_2}+\ldots
+u_{1\ldots n}e^{1\ldots n}
\label{U:decomp1}
\end{equation}
with coefficients $u,u_a,u_{a_1 a_2},\ldots,u_{1\ldots n}\in\F$,
which numbered by ordered multi-indices of length form $0$ to $n$.

Denote by $\cl^\F_k(p,q)$, ($k=0,\ldots n$) subspaces of the vector
space
$\cl^\F(p,q)$ that span over basis elements $e^{a_1\ldots
a_k}$. Elements of
$\cl^\F_k(p,q)$ are called {elements of rank $k$}.
Sometimes it is suitable to denote
$\ss{k}{U}\in\cl^\F_k(p,q)$.
We have
$$
\cl^\F(p,q)=\cl^\F_0(p,q)\oplus\ldots\oplus \cl^\F_n(p,q)
=\cl^\F_\even(p,q)\oplus\cl^\F_\odd(p,q),
$$
where
$$
\cl^\F_\even(p,q)=\cl^\F_0(p,q)\oplus\cl^\F_2(p,q)\oplus\ldots,
\quad
\cl^\F_\odd(p,q)=\cl^\F_1(p,q)\oplus\cl^\F_3(p,q)\oplus\ldots
$$
and
$$
\dimension \cl^\F_k(p,q)=C^k_n,\quad\dimension\cl^\F_\even(p,q)=
\dimension\cl^\F_\odd(p,q)=2^{n-1},
$$
$C^k_n$ are binomial coefficients. Let us take antisymmetric
coefficients
$u_{a_1\ldots a_k}=u_{[a_1\ldots
a_k]}\in\F$, where square brackets denote operation of alternation.
Consider an element
$$
\ss{k}{U}=\sum_{a_1<\cdots<a_k}u_{a_1\ldots a_k}e^{a_1\ldots a_k}\in\cl^\F_k(p,q).
$$
We have
$$
\ss{k}{U}=\sum_{a_1<\cdots<a_k}u_{a_1\ldots a_k}e^{a_1\ldots a_k}
=\frac{1}{k!}u_{b_1\ldots b_k}e^{b_1}\ldots e^{b_k},
$$
and
$$
U=ue+u_a e^a+\frac{1}{2!} u_{a_1 a_2}e^{a_1}e^{a_2}+\ldots
+\frac{1}{n!}u_{a_1\ldots a_n}e^{a_1}\ldots e^{a_n}.
$$
for $U$ from
(\ref{U:decomp1}).


\noindent{\bf The exterior product of Clifford algebra elements.}
Let us define
\begin{equation}
e^{i_1}\wedge e^{i_2}\ww e^{i_k}=e^{[i_1}e^{i_2}\ldots e^{i_k]}.
\label{cl:gr}
\end{equation}
In particular,
\begin{equation}
e^{i_1}\wedge e^{i_2}=\frac{1}{2}(e^{i_1}e^{i_2}-e^{i_2}e^{i_1})=
e^{i_1}e^{i_2}-\eta^{i_1 i_2}e,
\label{N2}
\end{equation}
\begin{eqnarray*}
e^{i_1}\wedge e^{i_2}\wedge e^{i_3}&=&
\frac{1}{6}(e^{i_1}e^{i_2}e^{i_3}+
e^{i_3}e^{i_1}e^{i_2}+e^{i_2}e^{i_3}e^{i_1}\\
&&-e^{i_2}e^{i_1}e^{i_3}-
e^{i_1}e^{i_3}e^{i_2}-e^{i_3}e^{i_2}e^{i_1})=\\
&=&e^{i_1}e^{i_2}e^{i_3}-\eta^{i_2 i_3}e^{i_1}+
\eta^{i_1 i_3}e^{i_2}-\eta^{i_1 i_2}e^{i_3},
\end{eqnarray*}
\begin{eqnarray*}
e^{i_1}\wedge e^{i_2}\wedge e^{i_3}\wedge e^{i_4}&=&
\frac{1}{24}(e^{i_1}e^{i_2}e^{i_3}e^{i_4}+\ldots)=\\
&=&e^{i_1}e^{i_2}e^{i_3}e^{i_4}-\eta^{i_3 i_4}e^{i_1}e^{i_2}
+\eta^{i_2 i_4}e^{i_1}e^{i_3}
-\eta^{i_2 i_3}e^{i_1}e^{i_4}\\
&&-\eta^{i_1 i_4}e^{i_2}e^{i_3}
+\eta^{i_1 i_3}e^{i_2}e^{i_4}-\eta^{i_1 i_2}e^{i_3}e^{i_4}\\
&&+(\eta^{i_1 i_4}\eta^{i_2 i_3}-\eta^{i_1 i_3}\eta^{i_2 i_4}+\eta^{i_1 i_2}\eta^{i_3 i_4})e
\end{eqnarray*}
From these formulas we get
\begin{eqnarray*}
e^{i_1}\wedge e^{i_2}=-e^{i_2}\wedge e^{i_1} \for i_1,i_2=1,\ldots,n;\\
e^{i_1}\ww e^{i_k}=e^{i_1}\ldots e^{i_k}=e^{i_1\ldots i_k} \for
i_1<\ldots<i_k.
\end{eqnarray*}

So, we arrive at the $2^n$ dimensional vector space
$\E$ with basis
(\ref{basis:cl}) and with two products of elements (the Clifford product
and the exterior product).


\section{Commutators and anti-commutators.}
There are well known formulas for the exterior product
\begin{equation}
\ss{k}{U}\wedge\ss{l}{V}=\ss{k+l}{W}
\label{wedge:formula}
\end{equation}
and for the Clifford product
\cite{Hestenes}
\begin{equation}
\ss{k}{U}\ss{l}{V}=\ss{|k-l|}{W}+\ss{|k-l|+2}{W}+\ldots+\ss{k+l}{W}.
\label{lemma:hes}
\end{equation}

Consider the commutator and the anti-commutator of Clifford algebra
elements
$$
[\ss{k}{U},\ss{l}{V}]=\ss{k}{U}\ss{l}{V}-\ss{l}{V}\ss{k}{U},\quad
\lbrace \ss{k}{U},\ss{l}{V}\rbrace =\ss{k}{U}\ss{l}{V}+\ss{l}{V}\ss{k}{U},\quad
\ss{k}{U}\ss{l}{V}=\frac{1}{2}[\ss{k}{U},\ss{l}{V}]+
\frac{1}{2}\lbrace \ss{k}{U},\ss{l}{V}\rbrace .
$$

\begin{theorem}\label{theorem100}. If
$\ss{k}{U}\in\cl^\F_k(p,q)$, $\ss{2}{V}\in\cl^\F_2(p,q)$,
then
$$
[\ss{k}{U},\ss{2}{V}]=\ss{k}{W}, \quad \hbox{for}\quad 1\leq k\leq n-1,
$$
where $n=p+q$.
\end{theorem}
\proof. \, We  must prove that
\begin{equation}
[e^{a_1\ldots a_k},e^{b_1 b_2}]\in\cl^\F_k(p,q).
\label{com:k2}
\end{equation}
If the multi-index $a_1\,\ldots\,a_k$ do not contain neither of
indices $b_1, b_2$, or contain both indices $b_1, b_2$, than the
commutator (\ref{com:k2}) is equal to zero. If the
$a_1\,\ldots\,a_k$ contain one of indices $b_1, b_2$, then
$$
e^{a_1\ldots a_k}e^{b_1 b_2}, e^{b_1 b_2} e^{a_1\ldots a_k}\in\cl^\F_k(p,q).
$$
\fin

\medskip

It follows from this theorem that the set
$\cl^\R_2(p,q)$ is closed with respect to the commutator and, hence, can
be considered as a Lie algebra. This Lie algebra is the real Lie algebra
of spinor Lie groups
$\Pin(p,q)$,
$\Pin_+(p,q)$,
$\Spin(p,q)$,
$\Spin_+(p,q)$ \cite{Lounesto}.

\begin{theorem}\label{theorem200}. Let $\ss{k}{U}, \ss{l}{V}, \ss{r}{W}$
be elements  of
$\cl^\F(p,q)$, $p+q=n$ of ranks $k,l,r$ respectively. Then
for all nonnegative integer $k\geq l$ the following formulas are valid
(let us remind that
$\ss{k}{W}=0$ for $k>n$ and for $k<0$):
$$
[\ss{k}{U},\ss{l}{V}]=\left\lbrace
\begin{array}{ll}
\ss{k-l+2}{W}+\ss{k-l+6}{W}+\ldots+\ss{k+l-2}{W}, & \mbox{\rm $l$ is even;}\\
\ss{k-l}{W}+\ss{k-l+4}{W}+\ldots+\ss{k+l-2}{W}, & \mbox{\rm $k$ is
even and $l$ is odd;}\\
\ss{k-l+2}{W}+\ss{k-l+6}{W}+\ldots+\ss{k+l}{W}, & \mbox{\rm $k,l$ are
odd.}
\end{array}
\right.
$$
$$
\lbrace \ss{k}{U},\ss{l}{V}\rbrace =\left\lbrace
\begin{array}{ll}
\ss{k-l}{W}+\ss{k-l+4}{W}+\ldots+\ss{k+l}{W}, & \mbox{\rm $l$ is even;}\\
\ss{k-l+2}{W}+\ss{k-l+6}{W}+\ldots+\ss{k+l}{W}, & \mbox{\rm $k$ is
even and $l$ is odd;}\\
\ss{k-l}{W}+\ss{k-l+4}{W}+\ldots+\ss{k+l-2}{W}, & \mbox{\rm  $k,l$ are
odd.}
\end{array}
\right.
$$
At the right hand parts of these formulas there are sums of elements
with different ranks and the increment between ranks is equal to $4$.

In particular, for $k=l$ we have formulas:
$$
[\ss{k}{U},\ss{k}{V}]=\left\lbrace
\begin{array}{ll}
\ss{2}{W}+\ss{6}{W}+\ldots+\ss{2k-2}{W}, & \mbox{\rm $k$ is
even;}\\
\ss{2}{W}+\ss{6}{W}+\ldots+\ss{2k}{W}, & \mbox{\rm $k$ is
odd.}
\end{array}
\right.
$$
$$
\lbrace \ss{k}{U},\ss{k}{V}\rbrace =\left\lbrace
\begin{array}{ll}
\ss{0}{W}+\ss{4}{W}+\ldots+\ss{2k}{W}, & \mbox{\rm $k$ is
even;}\\
\ss{0}{W}+\ss{4}{W}+\ldots+\ss{2k-2}{W}, & \mbox{\rm  $k$ is
odd.}
\end{array}
\right.
$$
\end{theorem}

Let us write down ranks of commutators and anti-commutators of
Clifford algebra elements of ranks from $1$ to $4$.
\begin{flalign*}
[\ss{1}{U},\ss{1}{V}]&=\ss{2}{W}&\qquad\lbrace \ss{1}{U},\ss{1}{V}\rbrace
&=\ss{0}{W}\\
[\ss{2}{U},\ss{1}{V}]&=\ss{1}{W}&\qquad\lbrace \ss{2}{U},\ss{1}{V}\rbrace
&=\ss{3}{W}\\
[\ss{2}{U},\ss{2}{V}]&=\ss{2}{W}&\qquad\lbrace \ss{2}{U},\ss{2}{V}\rbrace
&=\ss{0}{W}+\ss{4}{W}\\
[\ss{3}{U},\ss{1}{V}]&=\ss{4}{W}&\qquad\lbrace \ss{3}{U},\ss{1}{V}\rbrace
&=\ss{2}{W}\\
[\ss{3}{U},\ss{2}{V}]&=\ss{3}{W}&\qquad\lbrace \ss{3}{U},\ss{2}{V}\rbrace
&=\ss{1}{W}+\ss{5}{W}\\
[\ss{3}{U},\ss{3}{V}]&=\ss{2}{W}+\ss{6}{W}&\qquad\lbrace \ss{3}{U},\ss{3}{V}\rbrace
&=\ss{0}{W}+\ss{4}{W}\\
[\ss{4}{U},\ss{1}{V}]&=\ss{3}{W}&\qquad\lbrace \ss{4}{U},\ss{1}{V}\rbrace
&=\ss{5}{W}\\
[\ss{4}{U},\ss{2}{V}]&=\ss{4}{W}&\qquad\lbrace \ss{4}{U},\ss{2}{V}\rbrace
&=\ss{2}{W}+\ss{6}{W}\\
[\ss{4}{U},\ss{3}{V}]&=\ss{1}{W}+\ss{5}{W}&\qquad\lbrace \ss{4}{U},\ss{3}{V}\rbrace
&=\ss{3}{W}+\ss{7}{W}\\
[\ss{4}{U},\ss{4}{V}]&=\ss{2}{W}+\ss{6}{W}&\qquad\lbrace \ss{4}{U},\ss{4}{V}\rbrace
&=\ss{0}{W}+\ss{4}{W}+\ss{8}{W}\\
\end{flalign*}

\proof\,\, of Theorem 2. \, The formulas $e^a e^b+e^b
e^a=\eta^{ab}e$ lead to the following formulas:
\begin{eqnarray*}
[e^{a_1\ldots a_k}, e^{b_1\ldots b_l}] & = &
(1-(-1)^{kl-s})e^{a_1\ldots a_k}e^{b_1\ldots b_l},\\
\lbrace e^{a_1\ldots a_k}, e^{b_1\ldots b_l}\rbrace & = &
(1+(-1)^{kl-s})e^{a_1\ldots a_k}e^{b_1\ldots b_l},\\
e^{a_1\ldots a_k} e^{b_1\ldots b_l} &\in&\cl^\F_{k+l-2s}(p,q),
\end{eqnarray*}
where $s$ is a number of coincide indices in the ordered
multi-indices $a_1\,\ldots\,a_k$ and $b_1\,\ldots\,b_l$ and $0\leq
s\leq\min(k,l)$. From these formulas we get
$$
[e^{a_1\ldots a_k}, e^{b_1\ldots b_l}]=
\left\lbrace
\begin{array}{ll}
\cl_{k+l-2s}, & \mbox{if $kl-s$ is odd},\\
0,  & \mbox{if $kl-s$ is even}.
\end{array}
\right.
$$
$$
\lbrace e^{a_1\ldots a_k}, e^{b_1\ldots b_l}\rbrace=
\left\lbrace
\begin{array}{ll}
\cl_{k+l-2s}, & \mbox{if $kl-s$ is even},\\
0,  & \mbox{if $kl-s$ is odd}.
\end{array}
\right.
$$
Considering all possible variants of evenness of
$k,l,s$ in the last formulas, we conclude the proof of Theorem 2. \fin
\medskip

Let us consider in more details commutators and anti-commutators of
Clifford algebra elements for small dimensions $1\leq n\leq 5$.

We take a Clifford algebra element of rank $k$
$$
\ss{k}{U}=\frac{1}{k!}u_{a_1\ldots a_k}e^{a_1}\ww e^{a_k}\in\cl_k(p,q),
$$
where $u_{a_1\ldots a_k}=u_{[a_1\ldots a_k]}$ and $n=p+q$.
Let $\star\,:\,\cl_k(p,q)\to\cl_{n-k}(p,q)$ be the Hodge operation
$$
\star\ss{k}{U}=\frac{1}{k!(n-k)!}\varepsilon_{a_1\ldots a_n}
u^{a_1\ldots a_k}e^{a_{k+1}}\ww e^{a_n},
$$
where $\varepsilon_{a_1\ldots a_n}$ is a completely antisymmetric value,
$\varepsilon_{1\ldots n}=1$ and
$u^{a_1\ldots a_k}=\eta^{a_1 b_1}\ldots\eta^{a_k b_k}u_{b_1\ldots b_k}$.

Also, we need a bilinear operation
$\com\,:\,\cl_2^\F(p,q)\times\cl_2^\F(p,q)\to\cl_2^\F(p,q)$
\begin{eqnarray*}
&&\com(\frac{1}{2}u_{a_1a_2}e^{a_1}\wedge e^{a_2},
\frac{1}{2}v_{b_1b_2}e^{b_1}\wedge e^{b_2})\\
&=&
\frac{1}{2}u_{a_1a_2}v_{b_1b_2}(-\eta^{a_1 b_1}e^{a_2}\wedge e^{b_2}\\
&&-\eta^{a_2 b_2}e^{a_1}\wedge e^{b_1}
+\eta^{a_1 b_2}e^{a_2}\wedge e^{b_1}+\eta^{a_2
b_1}e^{a_1}\wedge e^{b_2}),
\end{eqnarray*}
where $u_{a_1a_2}=u_{[a_1a_2]}$,
$v_{b_1b_2}=v_{[b_1b_2]}$.
Evidently,
$\com(U,V)=-\com(V,U)$.
\medskip

Let $\varrho$ be the sign ($\pm 1$) of $\det\,\eta$.

\begin{theorem}\label{theorem300}. For $1\leq n\leq 5$ all commutators
$[\ss{k}{U},\ss{l}{V}]$ and all anti-commutators $\lbrace
\ss{k}{U},\ss{l}{V}\rbrace$ of Clifford algebra elements
$\ss{k}{U}\in\cl_k^\F(p,q)$, $\ss{k}{V}\in\cl_l^\F(p,q)$ can be
expressed with the aid of the exterior product, the Hodge $\star$
operation, and the $\Com$ operation:
\begin{flalign*}
n&=1&&\\
[\ss{1}{U},\ss{1}{V}]&=0&\qquad\lbrace \ss{1}{U},\ss{1}{V}\rbrace
&=2\star(\ss{1}{U}\wedge\star\ss{1}{V})\varrho,
\end{flalign*}
\begin{flalign*}
n&=2&&\\
[\ss{1}{U},\ss{1}{V}]&=2\ss{1}{U}\wedge\ss{1}{V}&\qquad\lbrace \ss{1}{U},\ss{1}{V}\rbrace &=2\star(\ss{1}{U}\wedge\star\ss{1}{V})\varrho\\
[\ss{1}{U},\ss{2}{V}]&=2\star(\ss{1}{U}\wedge\star\ss{2}{V})\varrho&\qquad\lbrace \ss{1}{U},\ss{2}{V}\rbrace &=0\\
[\ss{2}{U},\ss{1}{V}]&=-2\star(\star\ss{2}{U}\wedge\ss{1}{V})\varrho&\qquad\lbrace \ss{2}{U},\ss{1}{V}\rbrace &=0\\
[\ss{2}{U},\ss{2}{V}]&=0&\qquad\lbrace \ss{2}{U},\ss{2}{V}\rbrace
&=-2\star(\ss{2}{U}\wedge\star\ss{2}{V})\varrho,
\end{flalign*}
\begin{flalign*}
n&=3&&\\
[\ss{1}{U},\ss{1}{V}]&=2\ss{1}{U}\wedge\ss{1}{V}&\qquad\lbrace \ss{1}{U},\ss{1}{V}\rbrace &=2\star(\ss{1}{U}\wedge\star\ss{1}{V})\varrho\\
[\ss{1}{U},\ss{2}{V}]&=-2\star(\ss{1}{U}\wedge\star\ss{2}{V})\varrho&\qquad\lbrace \ss{1}{U},\ss{2}{V}\rbrace &=2\ss{1}{U}\wedge\ss{2}{V}\\
[\ss{1}{U},\ss{3}{V}]&=0&\qquad\lbrace \ss{1}{U},\ss{3}{V}\rbrace &=2\star(\ss{1}{U}\wedge\star\ss{3}{V})\varrho\\
[\ss{2}{U},\ss{1}{V}]&=-2\star(\star\ss{2}{U}\wedge\ss{1}{V})\varrho&\qquad\lbrace \ss{2}{U},\ss{1}{V}\rbrace &=2\ss{2}{U}\wedge\ss{1}{V}\\
[\ss{2}{U},\ss{2}{V}]&=-2\star\ss{1}{U}\wedge\star\ss{2}{V}\varrho&\qquad\lbrace \ss{2}{U},\ss{2}{V}\rbrace &=-2\star(\ss{2}{U}\wedge\star\ss{2}{V})\varrho\\
[\ss{2}{U},\ss{3}{V}]&=0&\qquad\lbrace \ss{2}{U},\ss{3}{V}\rbrace &=-2\star\ss{2}{U}\wedge\star\ss{3}{V}\varrho\\
[\ss{3}{U},\ss{1}{V}]&=0&\qquad\lbrace \ss{3}{U},\ss{1}{V}\rbrace &=2\star\ss{3}{U}\wedge\star\ss{1}{V}\varrho\\
[\ss{3}{U},\ss{2}{V}]&=0&\qquad\lbrace \ss{3}{U},\ss{2}{V}\rbrace &=-2\star\ss{3}{U}\wedge\star\ss{2}{V}\varrho\\
[\ss{3}{U},\ss{3}{V}]&=0&\qquad\lbrace \ss{3}{U},\ss{3}{V}\rbrace
&=-2\star\ss{3}{U}\wedge\star\ss{3}{V}\varrho,
\end{flalign*}
\begin{flalign*}
n&=4&&\\
[\ss{1}{U},\ss{1}{V}]&=2\ss{1}{U}\wedge\ss{1}{V}&\qquad\lbrace \ss{1}{U},\ss{1}{V}\rbrace &=2\star(\ss{1}{U}\wedge\star\ss{1}{V})\varrho\\
[\ss{1}{U},\ss{2}{V}]&=2\star(\ss{1}{U}\wedge\star\ss{2}{V})\varrho&\qquad\lbrace \ss{1}{U},\ss{2}{V}\rbrace &=2\ss{1}{U}\wedge\ss{2}{V}\\
[\ss{1}{U},\ss{3}{V}]&=2\ss{1}{U}\wedge\ss{3}{V}&\qquad\lbrace \ss{1}{U},\ss{3}{V}\rbrace &=2\star(\ss{1}{U}\wedge\star\ss{3}{V})\varrho\\
[\ss{1}{U},\ss{4}{V}]&=2\star(\ss{1}{U}\wedge\star\ss{4}{V})\varrho&\qquad\lbrace \ss{1}{U},\ss{4}{V}\rbrace &=0\\
[\ss{2}{U},\ss{1}{V}]&=-2\star(\star\ss{2}{U}\wedge\ss{1}{V})\varrho&\qquad\lbrace \ss{2}{U},\ss{1}{V}\rbrace &=2\ss{2}{U}\wedge\ss{1}{V}\\
[\ss{2}{U},\ss{2}{V}]&=\com(\ss{2}{U},\ss{2}{V})&\qquad\lbrace \ss{2}{U},\ss{2}{V}\rbrace &=2\ss{2}{U}\wedge\ss{2}{V}-2\star(\ss{2}{U}\wedge\star\ss{2}{V})\varrho\\
[\ss{2}{U},\ss{3}{V}]&=-2\star\ss{2}{U}\wedge\star\ss{3}{V}\varrho&\qquad\lbrace \ss{2}{U},\ss{3}{V}\rbrace &=2\star(\ss{2}{U}\wedge\star\ss{3}{V})\varrho\\
[\ss{2}{U},\ss{4}{V}]&=0&\qquad\lbrace \ss{2}{U},\ss{4}{V}\rbrace &=-2\star\ss{2}{U}\wedge\star\ss{4}{V}\varrho\\
[\ss{3}{U},\ss{1}{V}]&=2\ss{3}{U}\wedge\ss{1}{V}&\qquad\lbrace \ss{3}{U},\ss{1}{V}\rbrace &=-2\star(\star\ss{3}{U}\wedge\ss{1}{V})\varrho\\
[\ss{3}{U},\ss{2}{V}]&=2\star\ss{3}{U}\wedge\star\ss{2}{V}\varrho&\qquad\lbrace \ss{3}{U},\ss{2}{V}\rbrace &=2\star(\star\ss{2}{U}\wedge\ss{1}{V})\varrho\\
[\ss{3}{U},\ss{3}{V}]&=-2\star\ss{3}{U}\wedge\star\ss{3}{V}\varrho&\qquad\lbrace \ss{3}{U},\ss{3}{V}\rbrace &=-2\star(\ss{3}{U}\wedge\star\ss{3}{V})\varrho\\
[\ss{3}{U},\ss{4}{V}]&=-2\star\ss{3}{U}\wedge\star\ss{4}{V}\varrho&\qquad\lbrace \ss{3}{U},\ss{4}{V}\rbrace &=0\\
[\ss{4}{U},\ss{1}{V}]&=-2\star\ss{4}{U}\wedge\star\ss{1}{V}\varrho&\qquad\lbrace \ss{4}{U},\ss{1}{V}\rbrace &=0\\
[\ss{4}{U},\ss{2}{V}]&=0&\qquad\lbrace \ss{4}{U},\ss{2}{V}\rbrace &=-2\star\ss{4}{U}\wedge\star\ss{2}{V}\varrho\\
[\ss{4}{U},\ss{3}{V}]&=2\star\ss{4}{U}\wedge\star\ss{3}{V}\varrho&\qquad\lbrace \ss{4}{U},\ss{3}{V}\rbrace &=0\\
[\ss{4}{U},\ss{4}{V}]&=0&\qquad\lbrace \ss{4}{U},\ss{4}{V}\rbrace
&=2\star\ss{4}{U}\wedge\star\ss{4}{V}\varrho,
\end{flalign*}
\begin{flalign*}
n&=5&&\\
[\ss{1}{U},\ss{1}{V}]&=2\ss{1}{U}\wedge\ss{1}{V}&\qquad\lbrace \ss{1}{U},\ss{1}{V}\rbrace &=2\star(\ss{1}{U}\wedge\star\ss{1}{V})\varrho\\
[\ss{1}{U},\ss{2}{V}]&=-2\star(\ss{1}{U}\wedge\star\ss{2}{V})\varrho&\qquad\lbrace \ss{1}{U},\ss{2}{V}\rbrace &=2\ss{1}{U}\wedge\ss{2}{V}\\
[\ss{1}{U},\ss{3}{V}]&=2\ss{1}{U}\wedge\ss{3}{V}&\qquad\lbrace \ss{1}{U},\ss{3}{V}\rbrace &=2\star(\ss{1}{U}\wedge\star\ss{3}{V})\varrho\\
[\ss{1}{U},\ss{4}{V}]&=-2\star(\ss{1}{U}\wedge\star\ss{4}{V})\varrho&\qquad\lbrace \ss{1}{U},\ss{4}{V}\rbrace &=2\ss{1}{U}\wedge\ss{4}{V}\\
[\ss{1}{U},\ss{5}{V}]&=0&\qquad\lbrace \ss{1}{U},\ss{5}{V}\rbrace &=2\star(\ss{1}{U}\wedge\star\ss{5}{V})\varrho\\
[\ss{2}{U},\ss{1}{V}]&=-2\star(\star\ss{2}{U}\wedge\ss{1}{V})\varrho&\qquad\lbrace \ss{2}{U},\ss{1}{V}\rbrace &=2\ss{2}{U}\wedge\ss{1}{V}\\
[\ss{2}{U},\ss{2}{V}]&=\com(\ss{2}{U},\ss{2}{V})&\qquad\lbrace \ss{2}{U},\ss{2}{V}\rbrace &=2\ss{2}{U}\wedge\ss{2}{V}-2\star(\ss{2}{U}\wedge\star\ss{2}{V})\varrho\\
[\ss{2}{U},\ss{3}{V}]&=\star \com(\ss{2}{U},\star\ss{3}{V})&\qquad\lbrace \ss{2}{U},\ss{3}{V}\rbrace &=2\ss{2}{U}\wedge\ss{3}{V}-2\star(\ss{2}{U}\wedge\star\ss{3}{V})\varrho\\
[\ss{2}{U},\ss{4}{V}]&=-2\star\ss{2}{U}\wedge\star\ss{4}{V}\varrho&\qquad\lbrace \ss{2}{U},\ss{4}{V}\rbrace &=-2\star(\ss{2}{U}\wedge\star\ss{4}{V})\varrho\\
[\ss{2}{U},\ss{5}{V}]&=0&\qquad\lbrace \ss{2}{U},\ss{5}{V}\rbrace &=-2\star(\ss{2}{U}\wedge\star\ss{5}{V})\varrho\\
[\ss{3}{U},\ss{1}{V}]&=2\ss{3}{U}\wedge\ss{1}{V}&\qquad\lbrace \ss{3}{U},\ss{1}{V}\rbrace &=2\star(\star\ss{3}{U}\wedge\ss{1}{V})\varrho\\
[\ss{3}{U},\ss{2}{V}]&=\star \com(\star\ss{3}{U},\ss{2}{V})&\qquad\lbrace \ss{3}{U},\ss{2}{V}\rbrace &=2\ss{3}{U}\wedge\ss{2}{V}-2\star(\star\ss{3}{U}\wedge\ss{2}{V})\varrho\\
[\ss{3}{U},\ss{3}{V}]&=\com(\ss{3}{U},\ss{3}{V})&\qquad\lbrace \ss{3}{U},\ss{3}{V}\rbrace &=-2\star(\ss{3}{U}\wedge\star\ss{3}{V})\varrho+2\star\ss{3}{U}\wedge\star\ss{3}{V}\varrho\\
[\ss{3}{U},\ss{4}{V}]&=2\star(\ss{3}{U}\wedge\star\ss{4}{V})\varrho&\qquad\lbrace \ss{3}{U},\ss{4}{V}\rbrace &=-2\star\ss{3}{U}\wedge\star\ss{4}{V}\varrho\\
[\ss{3}{U},\ss{5}{V}]&=0&\qquad\lbrace \ss{3}{U},\ss{5}{V}\rbrace &=-2\star(\ss{3}{U}\wedge\star\ss{5}{V})\varrho\\
[\ss{4}{U},\ss{1}{V}]&=-2\star(\star\ss{4}{U}\wedge\ss{1}{V})\varrho&\qquad\lbrace \ss{4}{U},\ss{1}{V}\rbrace &=2\ss{4}{U}\wedge\ss{1}{V}\\
[\ss{4}{U},\ss{2}{V}]&=-2\star\ss{4}{U}\wedge\star\ss{2}{V}\varrho&\qquad\lbrace \ss{4}{U},\ss{2}{V}\rbrace &=-2\star(\star\ss{4}{U}\wedge\ss{2}{V})\varrho\\
[\ss{4}{U},\ss{3}{V}]&=-2\star(\star\ss{4}{U}\wedge\ss{3}{V})\varrho&\qquad\lbrace \ss{4}{U},\ss{3}{V}\rbrace &=-2\star\ss{4}{U}\wedge\star\ss{3}{V}\varrho\\
[\ss{4}{U},\ss{4}{V}]&=2\star\ss{4}{U}\wedge\star\ss{4}{V}\varrho&\qquad\lbrace \ss{4}{U},\ss{4}{V}\rbrace &=2\star(\ss{4}{U}\wedge\star\ss{4}{V})\varrho\\
[\ss{4}{U},\ss{5}{V}]&=0&\qquad\lbrace \ss{4}{U},\ss{5}{V}\rbrace &=2\star(\ss{4}{U}\wedge\star\ss{5}{V})\varrho\\
[\ss{5}{U},\ss{1}{V}]&=0&\qquad\lbrace \ss{5}{U},\ss{1}{V}\rbrace &=2\star(\star\ss{5}{U}\wedge\ss{1}{V})\varrho\\
[\ss{5}{U},\ss{2}{V}]&=0&\qquad\lbrace \ss{5}{U},\ss{2}{V}\rbrace &=-2\star(\star\ss{5}{U}\wedge\ss{2}{V})\varrho\\
[\ss{5}{U},\ss{3}{V}]&=0&\qquad\lbrace \ss{5}{U},\ss{3}{V}\rbrace &=-2\star(\star\ss{5}{U}\wedge\ss{3}{V})\varrho\\
[\ss{5}{U},\ss{4}{V}]&=0&\qquad\lbrace \ss{5}{U},\ss{4}{V}\rbrace &=2\star(\star\ss{5}{U}\wedge\ss{4}{V})\varrho\\
[\ss{5}{U},\ss{5}{V}]&=0&\qquad\lbrace \ss{5}{U},\ss{5}{V}\rbrace &=2\star(\ss{5}{U}\wedge\star\ss{5}{V})\varrho
\end{flalign*}
\end{theorem}

\proof\,. The proof is by direct calculation.\fin

Note that using Theorem \ref{theorem300} and the formula
$$
\ss{k}{U} \ss{l}{V} =
\frac{1}{2}[\ss{k}{U},\ss{l}{V}] +
\frac{1}{2}\lbrace \ss{k}{U},\ss{l}{V}\rbrace
$$
we can express a Clifford algebra elements product $\ss{k}{U}\ss{l}{V}$
via the exterior product, the Hodge $\star$ operation, and the $\Com$
operation.

\begin{theorem}\label{theorem400}. If $k$ is an integer $1\leq k\leq
n-1$ and
$[\ss{k}{U},\ss{2}{V}]=0$ for all
$\ss{k}{U}\in\cl_k^\F(p,q)$, then
$\ss{2}{V}=0$.
\end{theorem}

\proof. \, Assume that $\ss{2}{V}\neq0$. Let us prove that for every
$1\leq k\leq n-1$ there exists $\ss{k}{U}\in\cl_k(p,q)$ such that
$[\ss{k}{U},\ss{2}{V}]\neq0$. Let indices $l<m$ be such that
$v_{lm}\neq 0$ and $\ss{2}{V} = v_{lm}e^{lm}$, where $t=0$ or
$t=\sum_{r<s,(r,s)\neq(l,m}v_{rs}e^{rs}$. Consider an element $e^l
e^{a_1 \ldots a_{k-1}} \in\cl_k(p,q)$, $k>0$, where $\lbrace m,
l\rbrace  \cap \lbrace a_1 \ldots a_{k-1}\rbrace  = \emptyset$. We
have
\begin{eqnarray*}
[e^l e^{a_1 \ldots a_{k-1}}, e^{lm}] &=& e^l e^{a_1 \ldots a_{k-1}} e^{lm} -
e^{lm} e^l e^{a_1 \ldots a_{k-1}}\\ &=& (-1)^{k-1} \eta^{ll} e^{a_1 \ldots a_{k-1}} e^m +
\eta^{ll} e^m e^{a_1 \ldots a_{k-1}}\\ &=& (-1)^{k-1} (-1)^{k-1} \eta^{ll}
e^m e^{a_1 \ldots a_{k-1}} + \eta^{ll} e^m e^{a_1 \ldots a_{k-1}}\\ &=& 2\eta^{ll}
e^m e^{a_1 \ldots a_{k-1}} \neq 0.
\end{eqnarray*}
This is true for $k=1,\ldots,n-1$. Further,
\begin{eqnarray}
[\ss{k}{U},\ss{2}{V}] &=& [e^l e^{a_1 \ldots a_{k-1}}, v_{lm} e^{lm} + t]
\nonumber\\ &=&
[e^l e^{a_1 \ldots a_{k-1}}, v_{lm} e^{lm}] + [e^l e^{a_1 \ldots a_{k-1}},t]
\label{UkV2}\\
&=& v_{lm} [e^l e^{a_1 \ldots a_{k-1}}, e^{lm}] + [e^l e^{a_1 \ldots
a_{k-1}},t]. \nonumber
\end{eqnarray}
Let us prove that the first and the second summands at the right hand
part of this identity are linear independent and, hence,
$[\ss{k}{U},\ss{2}{V}]\neq 0$. The second summand is a sum of
commutators of the form $v_{ij}[e^l e^{a_1\ldots a_{k-1}},e^{ij}]$.
Suppose that $l\neq i$ and $l\neq j$. Then we get that terms
$[e^l e^{a_1\ldots a_{k-1}},e^{ij}]$ are linear independent with
$[e^l e^{a_1\ldots a_{k-1}},e^{lm}]=2\eta^{ll}e^m e^{a_1\ldots
a_{k-1}}$ as indices $l,m,a_1,\ldots,a_{k-1}$ are mutually different.

Without lost of generality, suppose that $i=l$ and $j=c$, where $0\leq
c\leq n$, $c\neq l$. Then the terms $[e^l e^{a_1\ldots a_{k-1}},e^{lm}]$
and $v_{ij}[e^l e^{a_1\ldots a_{k-1}},e^{ij}]$ are linear independent if
$e^l e^{a_1\ldots a_{k-1}} e^l e^c$ and $e^m e^{a_1\ldots a_{k-1}}$ are
linear independent. That means $c\neq m$ and the basis element $e^{lm}$
is already considered in the decomposition (\ref{UkV2}). Therefore we
have prove that the commutator (\ref{UkV2}) is not equal to zero. This
completes the proof of Theorem \ref{theorem400}.\fin


\section{Generators contraction formulas.}

\noindent{\bf A volume element}. For $\cl^\F(p,q)$, $p+q=n$ the basis
element of rank $n$ is called
{\it the volume element} and denoted by
$$
\ell=e^{1\ldots n}=e^1\ldots e^n=e^1\ww e^n=
\frac{1}{n!}\varepsilon_{a_1\ldots a_n}e^{a_1}\ldots e^{a_n}=
\frac{1}{n!}\varepsilon_{a_1\ldots a_n}e^{a_1}\ww e^{a_n},
$$
where $\varepsilon_{a_1\ldots a_n}$ is completely antisymmetric and
$\varepsilon_{1\ldots n}=1$.
We have formulas
\begin{eqnarray*}
\ell^2&=&(-1)^{\frac{n(n-1)}{2}}\det\,\eta\,e,\\
\ell^*&=&(-1)^{\frac{n(n-1)}{2}}\ell,\\
\ell\ss{k}{U}&=&(-1)^{k(n+1)}\ss{k}{U}\ell.
\end{eqnarray*}
If $n$ is odd, then the volume element $\ell$ commutes with all elements
of
$\cl^\F(p,q)$. If $n$ is even, then
$\ell$ commutes ($[\ell,U]=0$) with all even elements from
$\cl^\F_\even(p,q)$ and anticommutes ($\lbrace \ell,U\rbrace =0$) with
all odd elements from
$\cl^\F_\odd(p,q)$.

For even $n$ the center of algebra $\cl^\F(p,q)$ coincides with
$\cl^\F_0(p,q)$ and for odd $n$ the center of algebra $\cl^\F(p,q)$
coincides with $\cl^\F_0(p,q)\oplus\cl^\F_n(p,q)$.

Let $e^a$ be generators of $\cl(p,q)$, $p+q=n$. Denote
$e_a=\eta_{ab}e^b$.

\begin{theorem}\label{theorem500}. (Generators contraction formulas).
For any
$\ss{k}{U}\in\cl_k(p,q)$
\begin{equation}
e_a \ss{k}{U} e^a=e^a\ss{k}{U} e_a=(-1)^k(n-2k)\ss{k}{U}.
\label{ea:ea}
\end{equation}
\end{theorem}

\proof. \,
Let us prove that
$e_a e^{b_1}\ldots e^{b_k} e^a = (-1)^k (n-2k) e^{b_1}\ldots e^{b_k}$
for $b_1<\ldots<b_k$. We use the method of mathematical induction.
For $k=0$, using the relation $\eta_{ab}=\eta_{ba}$, we get
\begin{eqnarray*}
e_a e^a &=& e^b \eta_{ab} e^a = \frac{1}{2} \eta_{ab}
e^b e^a + \frac{1}{2} \eta_{ab} e^b e^a
\\ &=&
\frac{1}{2} \eta_{ab} (e^b e^a + e^a e^b)= n.
\end{eqnarray*}
Hence, the formula (\ref{ea:ea}) is valid for $k=0$. Suppose that
formula (\ref{ea:ea}) is valid for some $k>0$. Let us prove the
validity of formula (\ref{ea:ea}) for $k+1$. We have
\begin{eqnarray*}
e_a e^{b_1}\ldots e^{b_k} e^{b_{k+1}} e^a &=& e_a e^{b_1}\ldots e^{b_k}
(-e^a e^{b_{k+1}} + 2 \eta^{a b_{k+1}})\\ &=& -e_a e^{b_1}\ldots e^{b_k} e^a e^{b_{k+1}}
+ 2 \eta^{a b_{k+1}} e_a e^{b_1}\ldots e^{b_k}\\ &=& -(-1)^k (n-2k)
e^{b_1}\ldots e^{b_k} e^{b_{k+1}} + 2 e^{b_{k+1}} e^{b_1}\ldots e^{b_k}\\ &=&
(-1)^k (-n+2k+2) e^{b_1}\ldots e^{b_{k+1}}\\ &=& (-1)^{k+1} (n-2(k+1)) e^{b_1}\ldots e^{b_{k+1}}
\end{eqnarray*}
This completes the proof of Theorem \ref{theorem500}.\fin

Let us note some partial cases of formula (\ref{ea:ea}).
\begin{itemize}
\item if $n$ is even and $k=n/2$, then
$e_a \ss{k}{U} e^a=e^a\ss{k}{U} e_a=0$;
\item $e^a e_a=n$;
\item $e^a\ell e_a=(-1)^{n+1}n\ell$;
\item for $n=4$ we have
$$
e^a(\ss{0}{U}+\ss{1}{U}+\ss{2}{U}+\ss{3}{U}+\ss{4}{U})e_a=
4\ss{0}{U}-2\ss{1}{U}+2\ss{3}{U}-4\ss{4}{U}.
$$
\end{itemize}


\section{Conjugation operators in Clifford algebras}

\noindent{\bf Projection operators to vector subspaces
$\cl^\F_k(p,q)$}.
Suppose $U\in\cl^\F(p,q)$ is written in the form (\ref{U:decomp1}).
Then denote
$$
\langle U\rangle_k=\ss{k}{U}=\sum_{a_1<\cdots<a_k}u_{a_1\ldots
a_k}e^{a_1\ldots a_k}\in\cl^\F_k(p,q).
$$
From formulas (\ref{wedge:formula}),(\ref{lemma:hes}) we have
$$
\langle\ss{k}{U}\ss{l}{V}\rangle_{k+l}=\ss{k}{U}\wedge\ss{l}{V}.
$$
Using the projection operator to the one dimensional vector subspace
$\cl^\F_0$, we define an operation
$\Tr\,:\,\cl^\F\to\F$
$$
\Tr(U)=\langle U\rangle_0|_{e\to 1}.
$$
We say that $\Tr(U)$ is the {\it trace} of an element $U$. For example,
$$
\Tr(ue+u_a e^a +\ldots)=u.
$$
The following formulas give
the main property of the $\Tr$ operation
$$
\Tr(UV)=\Tr(VU),\quad \Tr([U,V])=0.
$$
These formulas follow from Theorem \ref{theorem200}.

\begin{theorem}\label{theorem600}. If an element $B\in\cl^\F(p,q)$
satisfies conditions
\begin{equation}
[B,e^a]=C^a,\quad a=1,\ldots,n
\label{B:com}
\end{equation}
for some given $C^a\in\cl^\F(p,q)$, $\Tr\,C^a=0$,$\,a=1,\ldots,n$,
then
\begin{eqnarray}
B&=&\sum_{k=1}^n\frac{1}{n+(-1)^{k+1}(n-2k)}\langle C^a e_a\rangle_k
+\alpha e\quad\hbox{для $n$ четного},
\label{B:formula}\\
B&=&\sum_{k=1}^{n-1}\frac{1}{n+(-1)^{k+1}(n-2k)}\langle C^a e_a\rangle_k
+\alpha e+\beta\ell\quad\hbox{для $n$ нечетного},
\nonumber
\end{eqnarray}
where $\alpha,\beta\in\F$ and $\ell=e^{1\ldots n}$.
\end{theorem}

In other words, formulas
(\ref{B:formula}) define $B$ up to a term from the center of
$\cl^\F(p,q)$.

\proof. Let us multiply left and right hand parts of
(\ref{B:com}) by $e_a$ and sum with respect to $a$. Then we get
$$
B e^a e_a-e^a B e_a=C^a e_a.
$$
Now we use formulas
(\ref{ea:ea})
\begin{eqnarray*}
B e^a e_a&=&\sum_{k=0}^n n\langle B\rangle_k,\\
e^a B e_a&=&\sum_{k=0}^n (-1)^k (n-2k)\langle B\rangle_k
\end{eqnarray*}
\fin

\medskip

\noindent{\bf Operations of conjugation}. Consider the following
operations of conjugation in $\cl^\F(p,q)$:
$$
U^{\wedge}=U|_{e^a\to-e^a},\quad
U^\sim=U|_{e^{a_1\ldots a_r}\to e^{a_r}\ldots e^{a_1}},\quad
\bar{U}=U|_{u_{a_1\ldots a_r}\to\bar{u}_{a_1\ldots a_r}}.
$$
In $\bar{u}_{a_1\ldots a_r}$ the bar means the complex conjugation.
The operation $U\to U^{\wedge}$ is called the {\it grade involution}.
The superposition of conjugations $U\to U^\sim$ and $U\to \bar{U}$
gives the {\it Clifford conjugation}
$U\to U^*=\bar{U}^\sim$. We have
\begin{eqnarray*}
U^{\wedge}&=\sum_{k=0}^n (-1)^k\langle U\rangle_k&=
\langle U\rangle_0-\langle U\rangle_1+\langle U\rangle_2-
\langle U\rangle_3+\langle U\rangle_4-\ldots,\\
U^\sim&=\sum_{k=0}^n (-1)^{\frac{k(k-1)}{2}} \langle U\rangle_k&=
\langle U\rangle_0+\langle U\rangle_1-\langle U\rangle_2-
\langle U\rangle_3+\langle U\rangle_4+\ldots
\end{eqnarray*}
and
$$
U^{\wedge\wedge}=U,\quad
U^{\sim\sim}=U,\quad
\bar{\bar{U}}=U,\quad
U^{**}=U,
$$
$$
(UV)^{\wedge}=U^{\wedge} V^{\wedge},\,
(\overline{UV})=\bar{U}\bar{V},\,
(UV)^\sim=V^\sim U^\sim,\,
(UV)^*=V^* U^*,
$$
$$
(U\wedge V)^{\wedge}=U^{\wedge}\wedge  V^{\wedge},\,
(\overline{U\wedge V})=\bar{U}\wedge \bar{V},\,
(U\wedge V)^\sim=V^\sim \wedge U^\sim,\,
(U\wedge V)^*=V^* \wedge U^*.
$$
We see that presented conjugation operations have the same properties
with respect to the Clifford multiplication and with respect to the
exterior multiplication.

Note that
$$
\Tr(U^*)=\overline{\Tr(U)}.
$$


\section{Unitary (Euclidean) spaces on Clifford algebras}
Denote $\cl^\F(n)=\cl^\F(n,0)$. Consider the following operation
$\cl^\F(n)\times\cl^\F(n)\to\F$
\begin{equation}
(U,V)=\Tr(U^*V).
\label{scal:mult}
\end{equation}
\begin{lem}. The operation $U,V\to(U,V)$ is a Hermitian (Euclidean)
scalar product of elements of
$\cl^\C(n)$ ($\cl^\R(n)$).
\end{lem}

\proof. \, We must prove that the properties
\begin{eqnarray}
&&(U,V)=\overline{(V,U)},\nonumber\\
&&(U,\lambda V)=\lambda(U,V),\nonumber\\
&&(U+V,W)=(U,W)+(V,W).\nonumber\\
&&(U,U)>0\for U\neq 0.
\label{herm:scal}
\end{eqnarray}
are valid for all
$U,V,W\in\cl^\F(n)$, $\lambda\in\F$.
The first three properties are evidently valid.
To prove
(\ref{herm:scal}) it is sufficient to prove that basis
(\ref{basis:cl}) is orthonormal with respect to the operation
$(\cdot,\cdot)$
$$
(e^{i_1\ldots i_k},e^{j_1\ldots j_l})=\left\lbrace
\begin{array}{ll}
0, & \mbox{if $(i_1\ldots i_k)\neq(j_1\ldots j_l)$};\\
1, & \mbox{if $(i_1\ldots i_k)=(j_1\ldots j_l)$}.
\end{array}
\right.
$$
If multi-indices $i_1\ldots i_k$ and $j_1\ldots j_l$ have $r$ common
indices, then
$$
e^{i_1\ldots i_k}e^{j_1\ldots j_l}\in\cl^\F_{k+l-2r}(n),
$$
i.e., $(e^{i_1\ldots i_k},e^{j_1\ldots j_l})=0$ for $k+l-2r>0$. We
have $k+l-2r=0$ iff multi-indices $i_1\ldots i_k$ and $j_1\ldots
j_l$ are identical. In this case
\begin{equation}
(e^{i_1\ldots i_k},e^{j_1\ldots j_l})=
\Tr(e^{i_k}\ldots e^{i_1}e^{i_1}\ldots e^{i_k})=\Tr(e)=1.
\label{orth:basis}
\end{equation}
Hence, basis
(\ref{basis:cl}) is orthonormal and for
$U\in\cl^\F(n)$ we have
\begin{equation}
(U,U)=\sum_{k=0}^n \sum_{a_1<\ldots<a_k}|u_{a_1\ldots a_k}|^2>0.
\label{scal:pos}
\end{equation}
This completes the proof of the Lemma.\fin

For Clifford algebras
$\cl^\F(p,q)$ with $q>0$ property (\ref{herm:scal}) is not valid.
In this case we define an operation
$\dagger : \cl^\F(p,q)\to\cl^\F(p,q)$
with the aid of the formulas
\begin{equation}
(e^{i_1\ldots i_k})^\dagger=e_{i_k}\ldots e_{i_1},\quad
\lambda^\dagger=\bar\lambda,
\label{hermconj}
\end{equation}
where $\lambda\in\C$ and
$e_a=\eta_{ab}e^b$.
We say that
$\dagger$ is the {\it operation of Hermitian conjugation} of Clifford
algebra elements.
It is easy to see that
$$
(U V)^\dagger=V^\dagger U^\dagger,\quad U^{\dagger\dagger}=U.
$$
Now we can define the Hermitian (Euclidean) scalar product of Clifford
algebra elements by the formula
$$
(U,V)=\Tr(U^\dagger V).
$$
In this case we have
$$
(e^{i_1\ldots i_k},e^{i_1\ldots i_k})=
\Tr(e_{i_k}\ldots e_{i_1}e^{i_1}\ldots e^{i_k})=\Tr(e)=1.
$$
(no summation w.r.t. $i_1,\ldots i_k$). Basis (\ref{basis:cl}) of
$\cl^\F(p,q)$ is orthonormal with respect to this scalar product
and property (\ref{scal:pos}) is valid. For generators $e^a$
formula (\ref{hermconj}) gives
\begin{eqnarray}
(e^a)^\dagger&=&e^a\for a=1,\ldots,p;
\label{ea:hermconj}\\
(e^a)^\dagger&=&-e^a\for a=p+1,\ldots,n.\nonumber
\end{eqnarray}

We present formulas (\ref{herm1}),(\ref{herm2}), which are equivalent to (\ref{hermconj}).

\medskip

\begin{theorem}\label{theorem700}. Let $p\geq 0$, $q\geq 0$, $n=p+q\geq 1$ be integer numbers.
Let us define the operation of Hermitian conjugation
$\dagger\,:\,\cl^\F_r(p,q)\to\cl^\F_r(p,q)$, $k=0,\ldots n$, with
the aid of the following formulas:
\begin{equation}
U^\dagger=\left\lbrace\begin{array}{lll}
U^* & \for (p,q)=(n,0);\\
-e^n U^{*\wedge} e^n & \for (p,q)=(n-1,1);\\
e^n e^{n-1} U^* e^{n-1} e^n & \for (p,q)=(n-2,2);\\
-e^n e^{n-1} e^{n-2} U^{*\wedge} e^{n-2} e^{n-1} e^n & \for (p,q)=(n-3,3);\\
\ldots\\
(-1)^q e^n \ldots e^1 U^{* \sharp} e^1 \ldots e^n & \for (p,q)=(0,n).
\end{array}
\right.
\label{herm1}
\end{equation}
If $q$ is odd, then
$\sharp$ is the operation of grade involution $\wedge$. Also, we may use the following
equivalent formulas:
\begin{equation}
U^\dagger=\left\lbrace\begin{array}{lll}
e^n \ldots e^1 U^{* \sharp} e^1 \ldots e^n & \for (p,q)=(n,0);\\
\ldots\\
e^3 e^2 e^1 U^{*} e^1 e^2 e^3 & \for (p,q)=(3,n-3);\\
e^2 e^1 U^{*\wedge} e^1 e^2 & \for (p,q)=(2,n-2);\\
e^1 U^* e^1 & \for (p,q)=(1,n-1);\\
U^{*\wedge} & \for (p,q)=(0,n).
\end{array}
\right.
\label{herm2}
\end{equation}
where $\sharp$ is the operation $\wedge$ for an even $p$. In this case
$$
U^{\dagger\dagger}=U,\quad (UV)^\dagger=V^\dagger U^\dagger,\quad
(U+V)^\dagger=U^\dagger+V^\dagger,
$$
$$
(\lambda U)^\dagger=\bar{\lambda}U^\dagger,\quad e^\dagger=e
\for U,V\in\cl^\F(p,q), \lambda\in\F
$$
and the operation
$(\cdot,\cdot):\cl^\F(p,q)\times\cl^\F(p,q)\to\F$
\begin{equation}
(U,V)=\Tr(U^\dagger V)
\label{herm:scal:prod}
\end{equation}
gives a Hermitian (Euclidian for $\F=\R$) scalar product in Clifford
algebra $\cl^\F(p,q)$.
\end{theorem}

\proof. \, It is sufficient to establish that formula (\ref{hermconj}) is equivalent
to the formula
$$
(e^{i_1\ldots i_k})^\dagger=
e^p \ldots e^1 (e^{i_1\ldots i_k})^{* \sharp} e^1 \ldots e^p,
$$
where $\sharp$ is $\wedge$ for even $p$ and to the formula
$$
(e^{i_1\ldots i_k})^\dagger=
(-1)^q e^n \ldots e^{p+1} (e^{i_1\ldots i_k})^{* \sharp} e^{p+1} \ldots e^n,
$$
where $\sharp$ is $\wedge$ for all odd $q$. Let $s$ be the number of common
elements in sets
$\lbrace i_1\ldots i_k \rbrace$ and $\lbrace 1 \ldots p \rbrace$. Using the identities
$\eta^{11}=\ldots=\eta^{pp}=1$,
$\eta^{p+1 p+1}=\ldots=\eta^{n n}=-1$, we transform the above formulas to the same form:
\begin{eqnarray*}
e^p \ldots e^1 (e^{i_1\ldots i_k})^{* \sharp} e^1 \ldots e^p &=&
e^p \ldots e^1 (e^{i_k} \ldots e^{i_1})^{\sharp} e^1 \ldots e^p \\ &=&
(-1)^{(p+1)k} e^p \ldots e^1 e^{i_k} \ldots e^{i_1} e^1 \ldots e^p \\ &=&
(-1)^{(p+1)k} (-1)^{kp-s} e^{i_k} \ldots e^{i_1}\\ &=&
(-1)^{k-s} e^{i_k} \ldots e^{i_1}.
\end{eqnarray*}

\begin{eqnarray*}
(-1)^{q} e^n \ldots e^{p+1} (e^{i_1\ldots i_k})^{* \sharp} e^{p+1} \ldots e^n &=&
(-1)^{q} e^n \ldots e^{p+1} (e^{i_k} \ldots e^{i_1})^{\sharp} e^{p+1} \ldots e^n \\ &=&
(-1)^q (-1)^{qk} e^n \ldots e^{p+1} e^{i_k} \ldots e^{i_1} e^{p+1} \ldots e^n \\ &=&
(-1)^q (-1)^{qk} (-1)^{kq-(k-s)} (-1)^q e^{i_k} \ldots e^{i_1}\\ &=&
(-1)^{k-s} e^{i_k} \ldots e^{i_1}.
\end{eqnarray*}

\begin{eqnarray*}
e_{i_k}\ldots e_{i_1} &=&
\eta_{i_1 i_1}\ldots \eta_{i_k i_k} e^{i^k}\ldots e^{i^1}\\ &=&
(-1)^{k-s} 1^s e^{i_k}\ldots e^{i_1}\\ &=&
(-1)^{k-s} e^{i_k}\ldots e^{i_1}
\end{eqnarray*}
(no summation over $i_1 \ldots i_k$). This completes the proof.\fin

Note that for the Hermitian scalar product (\ref{herm:scal:prod})
we have
$$
(AU,V)=(U,A^\dagger V)\quad \forall A,U,V\in\cl^\F(p,q).
$$
Consider an element $U\in\cl^\F(p,q)$.  If $U=U^\dagger$, then the
element $U$ is called {\it Hermitian}. If $U=-U^\dagger$, then the
element $U$ is called  {\it antiHermitian} Any element $\cl^\F(p,q)$
can be decomposed into the sum of Hermitian and antiHermitian
elements
$$
U=\frac{1}{2}(U+U^\dagger)+\frac{1}{2}(U-U^\dagger).
$$


\section{Hermitian idempotents and related structures}
In what follows we consider only complex Clifford algebras
$\cl(p,q)=\cl^\C(p,q)$. The element $t\in\cl(p,q)$ is said to be
{\it the Hermitian idempotent} if
$$
t^2=t,\quad t^\dagger=t.
$$
We say that two Hermitian idempotents $t$ and $\hat{t}$ are of the
same type, if there exists a unitary element $U\in\cl(p,q)$,
$U^\dagger=U^{-1}$ such that
$$
\hat{t}=U^{-1} t U.
$$
It can be shown that  for Clifford algebra $\cl(1,3)$ there exist
four types of Hermitian idempotents.

The set of clifford algebra elements
$$
I(t)=\{U\in\cl(p,q) : U=Ut\}
$$
is called {\it the left ideal} of Clifford algebra (generated by the
Hermitian idempotent $t$).

Let us define the set of Clifford algebra elements, which depend on
a Hermitian idempotent $t$
$$
K(t)=\{U\in\cl(p,q) : U=t U t\}.
$$
It is evident that
$K(t)\subseteq I(t)$. Note that $[U,t]=0$ for
$U\in K(t)$.

A left ideal that doesn't contain other left ideals except itself
and the trivial ideal (generated by $t=0$), is called {\it a minimal
left ideal}. A Hermitian idempotent, which generates a minimal left
ideal is called {\it primitive}. The main property of a left ideal
$I(t)$: if $U\in I(t)$ and $V\in\cl(p,q)$, then $V U\in I(t)$.

\medskip

The left ideal $I(t)$ is a vector space. The Hermitian scalar
product $U,V\in I(t)\to (U,V)=\Tr(U^\dagger V)$, gives us the
structure of unitary space on $I(t)$. Let us take an orthonormal
basis $\tau_1,\ldots\tau_d\in I(t)$, where $d=\dim\,I(t)$,
$\tau^l=\tau_l$
\begin{equation}
(\tau_k,\tau^l)=\delta_k^l,\quad k,l=1,\ldots,d.
\label{tau:basis}
\end{equation}
In the sequel we consider matrix representations of Clifford
algebra elements. Let $\Mat(d,\C)$ be the algebra of
$d$-dimensional matrices with complex elements. A matrix
$Q\in\Mat(d,\C)$ has elements $q^l_k$, $k,l=1,\ldots d$,
enumerated by two indices. The upper (first) index enumerates rows
of the matrix and the lower (second) index enumerates columns of
the matrix. The product $P=Q R$ of two matrices $Q=\|q^k_l\|,\,
R=\|r^k_l\|\in\Mat(d,\C)$ is defined by the usual formula
$$
p^k_m=q^k_l r^l_m,
$$
where at the right hand part we have summation over $l$ (from 1 to $d$). With
the aid of the basis
(\ref{tau:basis}) we may define three linear maps
\begin{eqnarray*}
\gamma &:& \cl(p,q)\to \Mat(d,\C),\\
\theta &:& K(t)\to\Mat(d,\C),\\
\rho &:& I(t)\to\C^d.
\end{eqnarray*}
Here the $d$-dimensional complex vector space
$\C^d$ is considered as the set of complex matrices with one column and $d$ raws.
We define the map $\gamma$ by the formula
\begin{equation}
U\tau_k=\gamma(U)^l_k\tau_l,
\label{mat:rep}
\end{equation}
where $U\in\cl(p,q)$  and $\gamma(U)=\|\gamma(U)_k^l\|\in\Mat(d,\C)$. Therefore,
\begin{equation}
\gamma(U)^k_l=(\tau^k,U\tau_l).
\label{mat:rep1}
\end{equation}
We claim that
$$
\gamma(U V)=\gamma(U)\gamma(V).
$$
Indeed,
$$
(U V)\tau_k=U(V\tau_k)=U\tau_l\gamma(V)_k^l=
\gamma(U)^m_l\gamma(V)^l_k\tau_m.
$$
Hence, we get a matrix representation of Clifford algebra elements. The dimension
of this representation is equal to the dimension of left ideal $I(t)$. A minimal left ideal
gives the matrix representation of Clifford algebra elements of the minimal dimension.

Denote
$$
\gamma^a=\gamma(e^a),\quad {\bf 1}=\gamma(e),
$$
where ${\bf 1}$ is the identity matrix. Relations for Clifford algebra generators
$e^a e^b+e^b e^a=\eta^{ab}e$ give relations for matrices
$$
\gamma^a \gamma^b + \gamma^b \gamma^a=\eta^{ab}{\bf 1}.
$$
Suppose that a representation
$\gamma : \cl(p,q)\to\Mat(d,\C)$ is generated by an orthonormal basis
$\tau_1,\ldots,\tau_d$ of a left ideal with the aid of formula
(\ref{mat:rep}). Then
\begin{equation}
\gamma(U^\dagger)=\gamma(U)^\dagger,\quad\forall U\in\cl(p,q),
\label{gamma:dagger}
\end{equation}
where $U^\dagger$ is the Hermitian conjugated element of Clifford
algebra $\cl(p,q)$ (see Theorem 7) and $\gamma(U)^\dagger$ is the
Hermitian conjugated matrix. To prove this fact we rewrite
(\ref{mat:rep1}) in the form
\begin{equation}
\gamma(U)_{kl}=(\tau_k,U\tau_l),
\label{mat:rep2}
\end{equation}
numerating elements of the matrix $\gamma(U)$ by two lower indices.
The operation of Hermitian scalar product
$$
(A,B)=\Tr(A^\dagger B), \quad A,B\in\cl(p,q)
$$
has properties
$$
(A,U B)=(U^\dagger A,B),\quad
(A,B)=\overline{(B,A)}.
$$
Therefore, from
(\ref{mat:rep2}) we get
$$
\gamma(U)_{kl}=(U^\dagger\tau_k,\tau_l),\quad
\overline{\gamma(U)}_{kl}=(\tau_l,U^\dagger\tau_k).
$$
Comparing the last formula with formula (\ref{mat:rep2}), we obtain
identity (\ref{gamma:dagger}).

Let us define the map
$\rho : I(t)\to\C^d$. If we take the decomposition of a left ideal element by the basis
$$
\Omega=\omega^k\tau_k\in I(t),
$$
then $\rho(\Omega)$ is the column
$$
\rho(\Omega)=(\omega^1\,\ldots\,\omega^d)^{\rm T},
$$
where $A^{\rm T}$ is the transposed matrix. In particular, we have
$$
\rho(\tau_k)=(0\ldots 1\ldots 0)^{\rm T}
$$
with only 1 on the $k$-th place of the column.

If $U\in\cl(p,q)$ and $\Omega\in I(t)$, then
$$
U\Omega=U\omega^n\tau_n=\omega^n\gamma(U)^k_n\tau_k\in I(t).
$$
That means
$$
\rho(U\Omega)=\gamma(U)\rho(\Omega).
$$

Let$ V\in K(t)$. Now we may define the map
$\theta : K(t)\to\Mat(d,\C)$ by the following formula:
$$
\tau_n V=\theta(V)^k_n \tau_k.
$$
From this formula we get
$$
\theta(V)^k_n=(\tau^k,\tau_n V).
$$
If $V\in K(t)$ and $\Omega\in I(t)$, then
$$
\Omega V=\omega^n \tau_n V=\omega^n \theta(V)^k_n\tau_k\in I(t).
$$
Therefore,
$$
\rho(\Omega V)=\theta(V)\rho(\Omega).
$$
If $U,V\in K(t)$, $\Omega\in I(t)$, then
$$
\rho(\Omega U V)=\theta(V)\theta(U)\rho(\Omega)=\theta(UV)\rho(\Omega).
$$
Thus we have
$$
\theta(UV)=\theta(V)\theta(U),
$$
i.e., at the right hand part we see multiplies in reverse order and so the map
$\theta$ in {\it an antirepresentation} of elements of $K(t)$.

If $U\in\cl(p,q)$, $V\in K(t)$, $\Omega\in I(t)$, then
то $U\Omega\in I(t)$, $\Omega V\in I(t)$ and
\begin{eqnarray*}
\rho(U\Omega V)&=&\gamma(U)\rho(\Omega V)=\gamma(U)\theta(V)\rho(\Omega),\\
\rho(U\Omega V)&=&\theta(V)\rho(U\Omega)
=\theta(V)\gamma(U)\rho(\Omega).
\end{eqnarray*}
It now follows that for all $U\in\cl(p,q)$, $V\in K(t)$
$$
[\gamma(U),\theta(V)]=0.
$$


\section{Normal representations of Clifford algebra elements}

Consider the Clifford algebra $\cl(p,q)$, $p+q=n$. It is well known
\cite{Lounesto}, that Clifford algebra elements can be represented by complex
matrices of the minimal dimension $2^{[\frac{n+1}{2}]}$. For odd $n=2k+1$
Clifford algebra elements can be represented by $2^{k+1}$ dimensional block-diagonal
complex matrices with two blocks of dimension $2^k$ on the diagonal (other elements are zero).

Let $\gamma : \cl(p,q)\to\Mat(d,\C)$ be a representation of Clifford algebra elements
that satisfy the following conditions:
\begin{itemize}
\item $d=2^{[\frac{n+1}{2}]}$.
\item $\gamma(U)^\dagger=\gamma(U^\dagger)$ for all $U\in\cl(p,q)$, where
$\gamma(U)^\dagger$ is the Hermitian conjugated matrix and
$U^\dagger$ is the Hermitian conjugated element of the Clifford
algebra (see Theorem 7).
\end{itemize}
Then $\gamma$ is called {\it a normal representation} of Clifford algebra elements.

In this section for any Clifford algebra $\cl(p,q)$, $p+q=n$ we give
some (standard) Hermitian idempotent $t$ and the corresponding
(standard) orthonormal basis of the left ideal $I(t)$. This basis
defines the normal representation for $\cl(p,q)$.

Consider generators $e^1,\ldots,e^p,e^{p+1},\ldots,e^n$ of $\cl(p,q)$, $p+q=n$ such that
$e^a e^b=-e^b e^a$ for $a\neq b$ and
$$
(e^a)^2=e\for a=1,\ldots,p;\quad
(e^a)^2=-e\for a=p+1,\ldots,n.
$$
For $n=1$ we take $t=e$ and for $n>1$ we take
\begin{equation}
t=\frac{1}{2}(e+i^a e^1)\prod_{k=1}^{[\frac{n}{2}]-1} \frac{1}{2}(e+i^{b_k}
e^{2k}e^{2k+1})\in\cl(p,q),
\label{t:idem}
\end{equation}
where
$$
a=\left\{
\begin{array}{ll}
0 & \for p\neq 0;\\
1 & \for p=0,
\end{array}
\right.
\quad
b_k=\left\{
\begin{array}{ll}
0 & \for 2k=p;\\
1 & \for 2k\neq p.
\end{array}
\right.
$$
In product (\ref{t:idem}) all terms are commute. Using formula
(\ref{ea:hermconj}), we get
$$
(\frac{1}{2}(e+i^a e^1))^2=\frac{1}{2}(e+i^a e^1)=(\frac{1}{2}(e+i^a
e^1))^\dagger,
$$
$$
(\frac{1}{2}(e+i^{b_k} e^{2k}e^{2k+1}))^2=
\frac{1}{2}(e+i^{b_k} e^{2k}e^{2k+1})=(\frac{1}{2}(e+i^{b_k}
e^{2k}e^{2k+1}))^\dagger.
$$
Therefore, $t$ is a Hermitian idempotent
$$
t^2=t,\quad t^\dagger=t.
$$
It can be shown that for even $n$ the idempotent $t$ is primitive.

Let us discuss some notations. If $e^1,\ldots,e^m$ are generators of the
Clifford algebra
$\cl(u,v)$, $u+v=m$, then for this Clifford algebra we may use the notation
\begin{equation}
\cl(e^1,\ldots,e^m)=\cl_\even(e^1,\ldots,e^m)\oplus
\cl_\odd(e^1,\ldots,e^m).
\label{cleee}
\end{equation}
The following elements:
\begin{equation}
e,e^a,e^{a_1}e^{a_2},\ldots, e^1\ldots e^n,\quad
a_1<a_2<\ldots
\label{clif:basis}
\end{equation}
are basis elements of $\cl(e^1,\ldots,e^n)$.

Now we may denote basis elements
(\ref{clif:basis}) by $c_k$, $k=1,\ldots,2^m$.

Consider the Clifford algebra
$$
Q=\cl(e^2,e^4,\ldots,e^n)\quad\hbox{for even $n$},
$$
$$
Q=\cl(e^2,e^4,\ldots,e^{n-1},e^n)\quad\hbox{for odd $n$}.
$$
The complex dimension of $Q$ is equal to $2^{\frac{n+1}{2}}$. Let
$c_k$, $k=1,\ldots,2^{\frac{n+1}{2}}$ be basis elements of the Clifford algebra $Q$.

For even $n$ the Clifford algebra $Q$ is defined by the set of generators $e^a$ with
even indices
$2\leq a\leq n$.
For odd $n$ the Clifford algebra $Q$ is defined by the set of generators $e^a$ that
consists of all generators with even indices
$2\leq a\leq n-1$ and one generator $e^n$. The Clifford algebra $Q$ is a subalgebra
of $\cl(p,q)$. The dimension of the algebra $Q$ is equal to  $2^{[(n+1)/2]}$.
Let us denote by $c_k$, $k=1,\ldots,2^{[(n+1)/2]}$ the basis elements of $Q$. Suppose that
in the sequence $c_k$ at first we take
$2^{[\frac{n+1}{2}]-1}$ even elements
\begin{equation}
e,e^2 e^4,\ldots
\label{even:basis}
\end{equation}
and at second we take
$2^{[\frac{n+1}{2}]-1}$ odd elements
\begin{equation}
e^2,e^4,\ldots
\label{odd:basis}
\end{equation}

\begin{theorem}\label{theorem800}. The following elements of the left ideal
$I(t)$:
$$
\tau_k=(\sqrt{2})^{[n/2]}c_k t,\quad
k=1,\ldots,2^{[\frac{n+1}{2}]}
$$
form an orthonormal basis of $I(t)$.
\end{theorem}

\proof. We have
$$
(\tau_k,\tau_l)=\Tr(\tau_k^\dagger\tau_l)=
(\sqrt{2})^{2[n/2]}\Tr(t^\dagger c_k^\dagger c_l t)=
(\sqrt{2})^{2[n/2]}\Tr(c_k^\dagger c_l t).
$$
Let us show that
$$
\left\lbrace\begin{array}{ll}
c_k^\dagger c_l=e & \for k=l;\\
\Tr(c_k^\dagger c_l t)=0 & \for k\neq l.
\end{array}
\right.
$$
If $c_k=e^{a_1}\ldots e^{a_r}$, then, according to formula
(\ref{hermconj}), $c_k^\dagger=e_{a_r}\ldots e_{a_1}$. Therefore,
$c_k^\dagger c_k=e$, $k=1,\ldots,2^{[(n+1)/2]}$ and
$$
(\tau_k,\tau_k)=(\sqrt{2})^{2[n/2]}\Tr\,t=1, \quad k=1,\ldots,2^{[(n+1)/2]}.
$$

Consider the case $k\neq l$. We see that
\begin{equation}
c_k^\dagger c_l=\pm e^{a_1}\ldots e^{a_s}\quad\hbox{for even $n$},
\label{cdaggerc}
\end{equation}
$$
c_k^\dagger c_l=\pm e^{a_1}\ldots e^{a_s},\quad\hbox{or}\quad
c_k^\dagger c_l=\pm e^{a_1}\ldots e^{a_s}e^n\quad
\hbox{for odd $n$},
$$
where $a_1<\ldots<a_s$ and $a_1,\ldots,a_s$ are even indices. The right hand part
of formula
(\ref{cdaggerc}) contains, at least, one multiplier and, hence,
$\Tr(c_k^\dagger c_l)=0$.

Let us write down the idempotent $t$ from (\ref{t:idem}) in the form
$$
t=2^{-[n/2]}e+\sum_{r=1}^n\sum_{b_1<\ldots<b_r}\lambda_{b_1\ldots b_r}
e^{b_1}\ldots e^{b_r},
$$
where $\lambda_{b_1\ldots b_r}\in C$ and every term
$\lambda_{b_1\ldots b_r}e^{b_1}\ldots e^{b_r}$ contains, at least, one generator $e^b$
with odd index (for odd $n$ the idempotent $t$ doesn't contain the generator $e^n$).
We get
$$
c_k^\dagger c_l t=2^{-[n/2]}c_k^\dagger c_l+
(\sum_{r=1}^n\sum_{b_1<\ldots<b_r}\lambda_{b_1\ldots b_r}c_k^\dagger c_l
e^{b_1}\ldots e^{b_r}).
$$
If we write the expression in brackets as a sum of the basis
elements of $\cl(e^1,\ldots,e^n)$, then every addend contains as a
multiplier, at least, one generator  $e^b$ with odd index. That
means the trace of every addend is equal to zero and
$$
\Tr(c_k^\dagger c_l t)=0,\for k\neq l.
$$
This completes the proof of the Theorem.\fin

Thus with the aid of the Hermitian idempotent $t$ and the orthonormal basis $\tau_k$ of
the left ideal $I(t)$ we give the normal representation $U\to\underline{U}$
of Clifford algebra elements with matrices form $\Mat(2^{[\frac{n+1}{2}]},\C)$.
This representation gives us possibility to transfer all main notions of the matrix algebra
to the Clifford algebra.

For $U\in\cl(p,q)$ the complex number
$$
\det\,U := \det\,\underline{U}\in\C.
$$
is called {\it the determinant} of $U$. It can be shown that
$\det\,U\in\R$ for $U\in\cl^\R(p,q)$. A complex number
$\lambda\in\C$ such that
$$
\det(U-\lambda e)=0
$$
is called {\it an eigen-value} of $U\in\cl(p,q)$. The set of all
eigen-value of an element $U\in\cl(p,q)$ is called {\it the
spectrum} of $U$. The spectrum of an element $U\in\cl(p,q)$ consists
of $2^{[\frac{n+1}{2}]}$ complex numbers. A Hermitian element
$U=U^\dagger\in\cl(p,q)$ has a real spectrum.

If $\lambda\in\C$ is an eigen-value of an element $U\in\cl(p,q)$ and
an element $V\in\cl(p,q)$, $V\neq 0$ satisfies the equality
$$
(U-\lambda e)V=0,
$$
then $V$ is called {a left eigen-element} of $U$. The left
eigen-element $V$ belongs to some left ideal of the Clifford algebra
$\cl(p,q)$ and this left ideal is not coincides with $\cl(p,q)$.
\medskip

Consider matrix representations of Clifford algebra generators for
small dimensions $n=p+q$.

For $n=2$ we take the idempotent $t=1/2(e+e^1)$
for the signatures $(p,q)=(2,0)$ and $(1,1)$. And we take
$t=1/2(e+ie^1)$
for the signature $(0,2)$. In this case we have the following basis
of the left ideal $I(t)$:
$\tau_1=\sqrt{2}et,
\tau_2=\sqrt{2}e^2t$.
This basis gives us the following matrix representations of Clifford algebra generators.

For $(p,q)=(2,0)$ we have
$$
\underline e^1=\left( \begin{array}{ll}
 1 & 0 \\
 0 & -1 \end{array}\right),\quad
\underline e^2=\left( \begin{array}{ll}
 0 & 1 \\
 1 & 0 \end{array}\right).
$$

For $ (p,q)=(1,1) $ we have
$$
\underline e^1=\left( \begin{array}{ll}
 1 & 0 \\
 0 & -1 \end{array}\right),\quad
\underline e^2=\left( \begin{array}{ll}
 0 & -1 \\
 1 & 0 \end{array}\right).
$$

For $ (p,q)=(0,2) $ we have
$$
\underline e^1=\left( \begin{array}{ll}
 -i & 0 \\
 0 & i \end{array}\right),\quad
\underline e^2=\left( \begin{array}{ll}
 0 & -1 \\
 1 & 0 \end{array}\right).
$$

For $n=4$ we take $t=1/2(e+e^1)1/2(e+ie^{23})$ for the signatures
(4,0), (1,3), (3,1);  $t=1/2(e+e^1)1/2(e+e^{23})$ for the signature
(2,2); $t=1/2(e+ie^1)1/2(e+ie^{23})$ for the signature
(0,4).
We have the following basis of the left ideal $I(t)$:
$$
\tau_1=2et,\quad \tau_2=2e^{24}t,\quad \tau_3=2e^2t,\quad
\tau_4=2e^4t.
$$
For $ (p,q)=(4,0) $ we have
$$
\underline e^1=\left( \begin{array}{llll}
 1 & 0 & 0 & 0\\
 0 & 1 & 0 & 0\\
 0 & 0 & -1 & 0\\
 0 & 0 & 0 & -1 \end{array}\right),\quad
\underline e^2=\left( \begin{array}{llll}
 0 & 0 & 1 & 0\\
 0 & 0 & 0 & 1\\
 1 & 0 & 0 & 0\\
 0 & 1 & 0 & 0 \end{array}\right),
$$
$$
\underline e^3=\left( \begin{array}{llll}
 0 & 0 & i & 0\\
 0 & 0 & 0 & -i\\
 -i & 0 & 0 & 0\\
 0 & i & 0 & 0 \end{array}\right),\quad
\underline e^4=\left( \begin{array}{llll}
 0 & 0 & 0 & 1\\
 0 & 0 & -1 & 0\\
 0 & -1 & 0 & 0\\
 1 & 0 & 0 & 0 \end{array}\right).
$$
For $ (p,q)=(3,1) $ we have
$$
\underline e^1=\left( \begin{array}{llll}
 1 & 0 & 0 & 0\\
 0 & 1 & 0 & 0\\
 0 & 0 & -1 & 0\\
 0 & 0 & 0 & -1 \end{array}\right),\quad
\underline e^2=\left( \begin{array}{llll}
 0 & 0 & 1 & 0\\
 0 & 0 & 0 & 1\\
 1 & 0 & 0 & 0\\
 0 & 1 & 0 & 0 \end{array}\right),
$$
$$
\underline e^3=\left( \begin{array}{llll}
 0 & 0 & i & 0\\
 0 & 0 & 0 & -i\\
 -i & 0 & 0 & 0\\
 0 & i & 0 & 0 \end{array}\right),\quad
\underline e^4=\left( \begin{array}{llll}
 0 & 0 & 0 & -1\\
 0 & 0 & -1 & 0\\
 0 & 1 & 0 & 0\\
 1 & 0 & 0 & 0 \end{array}\right).
$$
For $ (p,q)=(2,2) $ we have
$$
\underline e^1=\left( \begin{array}{llll}
 1 & 0 & 0 & 0\\
 0 & 1 & 0 & 0\\
 0 & 0 & -1 & 0\\
 0 & 0 & 0 & -1 \end{array}\right),\quad
\underline e^2=\left( \begin{array}{llll}
 0 & 0 & 1 & 0\\
 0 & 0 & 0 & 1\\
 1 & 0 & 0 & 0\\
 0 & 1 & 0 & 0 \end{array}\right),
$$
$$
\underline e^3=\left( \begin{array}{llll}
 0 & 0 & -1 & 0\\
 0 & 0 & 0 & 1\\
 1 & 0 & 0 & 0\\
 0 & -1 & 0 & 0 \end{array}\right),\quad
\underline e^4=\left( \begin{array}{llll}
 0 & 0 & 0 & -1\\
 0 & 0 & -1 & 0\\
 0 & 1 & 0 & 0\\
 1 & 0 & 0 & 0 \end{array}\right).
$$
For $ (p,q)=(1,3) $ we have
$$
\underline e^1=\left( \begin{array}{llll}
 1 & 0 & 0 & 0\\
 0 & 1 & 0 & 0\\
 0 & 0 & -1 & 0\\
 0 & 0 & 0 & -1 \end{array}\right),\quad
\underline e^2=\left( \begin{array}{llll}
 0 & 0 & -1 & 0\\
 0 & 0 & 0 & 1\\
 1 & 0 & 0 & 0\\
 0 & -1 & 0 & 0 \end{array}\right),
$$
$$
\underline e^3=\left( \begin{array}{llll}
 0 & 0 & i & 0\\
 0 & 0 & 0 & i\\
 i & 0 & 0 & 0\\
 0 & i & 0 & 0 \end{array}\right),\quad
\underline e^4=\left( \begin{array}{llll}
 0 & 0 & 0 & -1\\
 0 & 0 & -1 & 0\\
 0 & 1 & 0 & 0\\
 1 & 0 & 0 & 0 \end{array}\right).
$$
For $ (p,q)=(0,4) $ we have
$$
\underline e^1=\left( \begin{array}{llll}
 -i & 0 & 0 & 0\\
 0 & -i & 0 & 0\\
 0 & 0 & i & 0\\
 0 & 0 & 0 & i \end{array}\right),\quad
\underline e^2=\left( \begin{array}{llll}
 0 & 0 & -1 & 0\\
 0 & 0 & 0 & 1\\
 1 & 0 & 0 & 0\\
 0 & -1 & 0 & 0 \end{array}\right),
$$
$$
\underline e^3=\left( \begin{array}{llll}
 0 & 0 & i & 0\\
 0 & 0 & 0 & i\\
 i & 0 & 0 & 0\\
 0 & i & 0 & 0 \end{array}\right),\quad
\underline e^4=\left( \begin{array}{llll}
 0 & 0 & 0 & -1\\
 0 & 0 & -1 & 0\\
 0 & 1 & 0 & 0\\
 1 & 0 & 0 & 0 \end{array}\right).
$$

If for the case $(p,q)=(1,3)$ we take the following basis of the
left ideal $I(t)$: $\tau_1=-2et, \tau_2=2e^{24}t, \tau_3=2e^4t,
\tau_4=2e^2t$, then we get the well known Dirac representation of
generators
$$
\underline e^1=\left( \begin{array}{llll}
 1 & 0 & 0 & 0\\
 0 & 1 & 0 & 0\\
 0 & 0 & -1 & 0\\
 0 & 0 & 0 & -1 \end{array}\right),\quad
\underline e^2=\left( \begin{array}{llll}
 0 & 0 & 0 & 1\\
 0 & 0 & 1 & 0\\
 0 & -1 & 0 & 0\\
 -1 & 0 & 0 & 0 \end{array}\right),
$$
$$
\underline e^3=\left( \begin{array}{llll}
 0 & 0 & 0 & -i\\
 0 & 0 & i & 0\\
 0 & i & 0 & 0\\
 -i & 0 & 0 & 0 \end{array}\right),\quad
\underline e^4=\left( \begin{array}{llll}
 0 & 0 & 1 & 0\\
 0 & 0 & 0 & -1\\
 -1 & 0 & 0 & 0\\
 0 & 1 & 0 & 0 \end{array}\right).
$$

Now let us consider matrix representations of Clifford algebra
generators for odd $n=p+q=1,3,5$. We get block-diagonal matrices.

For $ (p,q)=(1,0)$ we have
$$
 t=e, \tau_1=(1/\sqrt{2})(e+e^1)t, \tau_2=(1/\sqrt{2})(e-e^1)t
$$
and
$$
\underline e^1=\left( \begin{array}{ll}
 1 & 0 \\
 0 & -1 \end{array}\right).
$$

For
$(p,q)=(0,1)$ we have
$$
t=e, \tau_1=(1/\sqrt{2})(e-ie^1)t, \tau_2=(1/\sqrt{2})(e+ie^1)t
$$
and
$$
\underline e^1=\left( \begin{array}{ll}
 i & 0 \\
 0 & -i \end{array}\right).
$$

For
$ (p,q)=(3,0)$ we have
$$
t=1/2(e+e^1), \tau_1=(e-ie^{23})t, \tau_2=(e^2-ie^3)t, \tau_3=(e^2+ie^3)t,
\tau_4=(e+ie^{23})t
$$
and
$$
\underline e^1=\left( \begin{array}{llll}
 1 & 0 & 0 & 0\\
 0 & -1 & 0 & 0\\
 0 & 0 & -1 & 0\\
 0 & 0 & 0 & 1 \end{array}\right),\quad
\underline e^2=\left( \begin{array}{llll}
 0 & 1 & 0 & 0\\
 1 & 0 & 0 & 0\\
 0 & 0 & 0 & 1\\
 0 & 0 & 1 & 0 \end{array}\right),\quad
\underline e^3=\left( \begin{array}{llll}
 0 & -i & 0 & 0\\
 i & 0 & 0 & 0\\
 0 & 0 & 0 & -i\\
 0 & 0 & i & 0 \end{array}\right).
$$

For $ (p,q)=(2,1)$ we have
$$
t=1/2(e+e^1), \tau_1=(e+e^{23})t, \tau_2=(e^2+e^3)t, \tau_3=(e^2-e^3)t,
\tau_4=(e-e^{23})t
$$
and
$$
\underline e^1=\left( \begin{array}{llll}
 1 & 0 & 0 & 0\\
 0 & -1 & 0 & 0\\
 0 & 0 & -1 & 0\\
 0 & 0 & 0 & 1 \end{array}\right),\quad
\underline e^2=\left( \begin{array}{llll}
 0 & 1 & 0 & 0\\
 1 & 0 & 0 & 0\\
 0 & 0 & 0 & 1\\
 0 & 0 & 1 & 0 \end{array}\right),\quad
\underline e^3=\left( \begin{array}{llll}
 0 & -1 & 0 & 0\\
 1 & 0 & 0 & 0\\
 0 & 0 & 0 & -1\\
 0 & 0 & 1 & 0 \end{array}\right).
$$

For
$ (p,q)=(1,2)$ we have
$$
t=1/2(e+e^1), \tau_1=(e-ie^{23})t, \tau_2=(e^2+ie^3)t, \tau_3=(e^2-ie^3)t,
\tau_4=(e+ie^{23})t
$$
and
$$
\underline e^1=\left( \begin{array}{llll}
 1 & 0 & 0 & 0\\
 0 & -1 & 0 & 0\\
 0 & 0 & -1 & 0\\
 0 & 0 & 0 & 1 \end{array}\right),\quad
\underline e^2=\left( \begin{array}{llll}
 0 & -1 & 0 & 0\\
 1 & 0 & 0 & 0\\
 0 & 0 & 0 & 1\\
 0 & 0 & -1 & 0 \end{array}\right),\quad
\underline e^3=\left( \begin{array}{llll}
 0 & -i & 0 & 0\\
 -i & 0 & 0 & 0\\
 0 & 0 & 0 & i\\
 0 & 0 & i & 0 \end{array}\right).
$$

For
$ (p,q)=(0,3)$ we have
$$
t=1/2(e+ie^1), \tau_1=(e-ie^{23})t, \tau_2=(e^2+ie^3)t, \tau_3=(e^2-ie^3)t,
\tau_4=(e+ie^{23})t
$$
and
$$
\underline e^1=\left( \begin{array}{llll}
 -i & 0 & 0 & 0\\
 0 & i & 0 & 0\\
 0 & 0 & i & 0\\
 0 & 0 & 0 & -i \end{array}\right),\quad
\underline e^2=\left( \begin{array}{llll}
 0 & -1 & 0 & 0\\
 1 & 0 & 0 & 0\\
 0 & 0 & 0 & 1\\
 0 & 0 & -1 & 0 \end{array}\right),\quad
\underline e^3=\left( \begin{array}{llll}
 0 & -i & 0 & 0\\
 -i & 0 & 0 & 0\\
 0 & 0 & 0 & i\\
 0 & 0 & i & 0 \end{array}\right).
$$

For
$ (p,q)=(5,0)$ we have
$$
t=1/2(e+e^1)1/2(e+ie^{23}, \tau_1=\sqrt{2}(e-ie^{45})t, \tau_2=\sqrt{2}(e^{24}-ie^{25})t,
\tau_3=\sqrt{2}(e^2-ie^{245})t,
$$
$$
\tau_4=\sqrt{2}(e^4-ie^{5})t,
\tau_5=\sqrt{2}(e^4+ie^{5})t,
\tau_6=\sqrt{2}(e^2+ie^{245})t,
$$
$$
 \tau_7=\sqrt{2}(e^{24}+ie^{25})t,
\tau_8=\sqrt{2}(e+ie^{45})t
$$
and
$$
\underline e^1=\left( \begin{array}{llllllll}
 1 & 0 & 0 & 0 & 0 & 0 & 0 & 0\\
 0 & 1 & 0 & 0 & 0 & 0 & 0 & 0\\
 0 & 0 & -1 & 0 & 0 & 0 & 0 & 0\\
 0 & 0 & 0 & -1 & 0 & 0 & 0 & 0\\
 0 & 0 & 0 & 0 & -1 & 0 & 0 & 0\\
 0 & 0 & 0 & 0 & 0 & -1 & 0 & 0\\
 0 & 0 & 0 & 0 & 0 & 0 & 1 & 0\\
 0 & 0 & 0 & 0 & 0 & 0 & 0 & 1 \end{array}\right),\quad
\underline e^2=\left( \begin{array}{llllllll}
 0 & 0 & 1 & 0 & 0 & 0 & 0 & 0\\
 0 & 0 & 0 & 1 & 0 & 0 & 0 & 0\\
 1 & 0 & 0 & 0 & 0 & 0 & 0 & 0\\
 0 & 1 & 0 & 0 & 0 & 0 & 0 & 0\\
 0 & 0 & 0 & 0 & 0 & 0 & 1 & 0\\
 0 & 0 & 0 & 0 & 0 & 0 & 0 & 1\\
 0 & 0 & 0 & 0 & 1 & 0 & 0 & 0\\
 0 & 0 & 0 & 0 & 0 & 1 & 0 & 0 \end{array}\right),
$$
$$
\underline e^3=\left( \begin{array}{llllllll}
 0 & 0 & i & 0 & 0 & 0 & 0 & 0\\
 0 & 0 & 0 & -i & 0 & 0 & 0 & 0\\
 -i & 0 & 0 & 0 & 0 & 0 & 0 & 0\\
 0 & i & 0 & 0 & 0 & 0 & 0 & 0\\
 0 & 0 & 0 & 0 & 0 & 0 & i & 0\\
 0 & 0 & 0 & 0 & 0 & 0 & 0 & -i\\
 0 & 0 & 0 & 0 & -i & 0 & 0 & 0\\
 0 & 0 & 0 & 0 & 0 & i & 0 & 0 \end{array}\right),\quad
\underline e^4=\left( \begin{array}{llllllll}
 0 & 0 & 0 & 1 & 0 & 0 & 0 & 0\\
 0 & 0 & -1 & 0 & 0 & 0 & 0 & 0\\
 0 & -1 & 0 & 0 & 0 & 0 & 0 & 0\\
 1 & 0 & 0 & 0 & 0 & 0 & 0 & 0\\
 0 & 0 & 0 & 0 & 0 & 0 & 0 & 1\\
 0 & 0 & 0 & 0 & 0 & 0 & -1 & 0\\
 0 & 0 & 0 & 0 & 0 & -1 & 0 & 0\\
 0 & 0 & 0 & 0 & 1 & 0 & 0 & 0 \end{array}\right),
$$
$$
\underline e^5=\left( \begin{array}{llllllll}
 0 & 0 & 0 & -i & 0 & 0 & 0 & 0\\
 0 & 0 & -i & 0 & 0 & 0 & 0 & 0\\
 0 & i & 0 & 0 & 0 & 0 & 0 & 0\\
 i & 0 & 0 & 0 & 0 & 0 & 0 & 0\\
 0 & 0 & 0 & 0 & 0 & 0 & 0 & -i\\
 0 & 0 & 0 & 0 & 0 & 0 & -i & 0\\
 0 & 0 & 0 & 0 & 0 & i & 0 & 0\\
 0 & 0 & 0 & 0 & i & 0 & 0 & 0 \end{array}\right).
$$

For
$ (p,q)=(4,1)$ we have
$$
t=1/2(e+e^1)1/2(e+ie^{23}, \tau_1=\sqrt{2}(e-e^{45})t, \tau_2=\sqrt{2}(e^{24}-e^{25})t,
\tau_3=\sqrt{2}(e^2-e^{245})t,
$$
$$
 \tau_4=\sqrt{2}(e^4-e^{5})t, \tau_5=\sqrt{2}(e^4+e^{5})t,
\tau_6=\sqrt{2}(e^2+e^{245})t,
$$
$$
\tau_7=\sqrt{2}(e^{24}+e^{25})t,
\tau_8=\sqrt{2}(e+e^{45})t.
$$
In this case the first four generators have the same representation as in the previous
case of signature $(5,0)$ and the last generator has the following representation
$$
\underline e^5=\left( \begin{array}{llllllll}
 0 & 0 & 0 & 1 & 0 & 0 & 0 & 0\\
 0 & 0 & 1 & 0 & 0 & 0 & 0 & 0\\
 0 & -1 & 0 & 0 & 0 & 0 & 0 & 0\\
 -1 & 0 & 0 & 0 & 0 & 0 & 0 & 0\\
 0 & 0 & 0 & 0 & 0 & 0 & 0 & 1\\
 0 & 0 & 0 & 0 & 0 & 0 & 1 & 0\\
 0 & 0 & 0 & 0 & 0 & -1 & 0 & 0\\
 0 & 0 & 0 & 0 & -1 & 0 & 0 & 0 \end{array}\right).
$$

For
$ (p,q)=(3,2)$ we have
$$
t=1/2(e+e^1)1/2(e+ie^{23}), \tau_1=\sqrt{2}(e-ie^{45})t, \tau_2=\sqrt{2}(e^{2}-ie^{245})t,
\tau_3=\sqrt{2}(e^4+ie^5)t,
$$
$$
\tau_4=\sqrt{2}(e^{24}+ie^{25})t, \tau_5=\sqrt{2}(e^{24}-ie^{25})t,
\tau_6=\sqrt{2}(e^4-ie^{5})t,
$$
$$
\tau_7=\sqrt{2}(e^{2}+ie^{245})t,
\tau_8=\sqrt{2}(e+ie^{45})t
$$
and
$$
\underline e^1=\left( \begin{array}{llllllll}
 1 & 0 & 0 & 0 & 0 & 0 & 0 & 0\\
 0 & -1 & 0 & 0 & 0 & 0 & 0 & 0\\
 0 & 0 & -1 & 0 & 0 & 0 & 0 & 0\\
 0 & 0 & 0 & 1 & 0 & 0 & 0 & 0\\
 0 & 0 & 0 & 0 & 1 & 0 & 0 & 0\\
 0 & 0 & 0 & 0 & 0 & -1 & 0 & 0\\
 0 & 0 & 0 & 0 & 0 & 0 & -1 & 0\\
 0 & 0 & 0 & 0 & 0 & 0 & 0 & 1 \end{array}\right),\quad
\underline e^2=\left( \begin{array}{llllllll}
 0 & 1 & 0 & 0 & 0 & 0 & 0 & 0\\
 1 & 0 & 0 & 0 & 0 & 0 & 0 & 0\\
 0 & 0 & 0 & 1 & 0 & 0 & 0 & 0\\
 0 & 0 & 1 & 0 & 0 & 0 & 0 & 0\\
 0 & 0 & 0 & 0 & 0 & 1 & 0 & 0\\
 0 & 0 & 0 & 0 & 1 & 0 & 0 & 0\\
 0 & 0 & 0 & 0 & 0 & 0 & 0 & 1\\
 0 & 0 & 0 & 0 & 0 & 0 & 1 & 0 \end{array}\right),
$$
$$
\underline e^3=\left( \begin{array}{llllllll}
 0 & i & 0 & 0 & 0 & 0 & 0 & 0\\
 -i & 0 & 0 & 0 & 0 & 0 & 0 & 0\\
 0 & 0 & 0 & i & 0 & 0 & 0 & 0\\
 0 & 0 & -i & 0 & 0 & 0 & 0 & 0\\
 0 & 0 & 0 & 0 & 0 & -i & 0 & 0\\
 0 & 0 & 0 & 0 & i & 0 & 0 & 0\\
 0 & 0 & 0 & 0 & 0 & 0 & 0 & -i\\
 0 & 0 & 0 & 0 & 0 & 0 & i & 0 \end{array}\right),\quad
\underline e^4=\left( \begin{array}{llllllll}
 0 & 0 & -1 & 0 & 0 & 0 & 0 & 0\\
 0 & 0 & 0 & 1 & 0 & 0 & 0 & 0\\
 1 & 0 & 0 & 0 & 0 & 0 & 0 & 0\\
 0 & -1 & 0 & 0 & 0 & 0 & 0 & 0\\
 0 & 0 & 0 & 0 & 0 & 0 & -1 & 0\\
 0 & 0 & 0 & 0 & 0 & 0 & 0 & 1\\
 0 & 0 & 0 & 0 & 1 & 0 & 0 & 0\\
 0 & 0 & 0 & 0 & 0 & -1 & 0 & 0 \end{array}\right),
$$
$$
\underline e^5=\left( \begin{array}{llllllll}
 0 & 0 & -i & 0 & 0 & 0 & 0 & 0\\
 0 & 0 & 0 & i & 0 & 0 & 0 & 0\\
 -i & 0 & 0 & 0 & 0 & 0 & 0 & 0\\
 0 & i & 0 & 0 & 0 & 0 & 0 & 0\\
 0 & 0 & 0 & 0 & 0 & 0 & -i & 0\\
 0 & 0 & 0 & 0 & 0 & 0 & 0 & i\\
 0 & 0 & 0 & 0 & -i & 0 & 0 & 0\\
 0 & 0 & 0 & 0 & 0 & i & 0 & 0 \end{array}\right).
$$

For $ (p,q)=(2,3)$ we take $t=1/2(e+e^1)1/2(e+e^{23})$ and we have
the same basis as in the case of signature $(3,2)$.
Representations of all generators, except $e^3$, have the same
form as for the signature $(3,2)$ and
$$
\underline e^3=\left( \begin{array}{llllllll}
 0 & -1 & 0 & 0 & 0 & 0 & 0 & 0\\
 1 & 0 & 0 & 0 & 0 & 0 & 0 & 0\\
 0 & 0 & 0 & -1 & 0 & 0 & 0 & 0\\
 0 & 0 & 1 & 0 & 0 & 0 & 0 & 0\\
 0 & 0 & 0 & 0 & 0 & 1 & 0 & 0\\
 0 & 0 & 0 & 0 & -1 & 0 & 0 & 0\\
 0 & 0 & 0 & 0 & 0 & 0 & 0 & 1\\
 0 & 0 & 0 & 0 & 0 & 0 & -1 & 0 \end{array}\right).
$$

For
$ (p,q)=(1,4)$ we take
$t=1/2(e+e^1)1/2(e+ie^{23})$. We have the same basis as for $(3,2)$ and
$$
\underline e^1=\left( \begin{array}{llllllll}
 1 & 0 & 0 & 0 & 0 & 0 & 0 & 0\\
 0 & -1 & 0 & 0 & 0 & 0 & 0 & 0\\
 0 & 0 & -1 & 0 & 0 & 0 & 0 & 0\\
 0 & 0 & 0 & 1 & 0 & 0 & 0 & 0\\
 0 & 0 & 0 & 0 & 1 & 0 & 0 & 0\\
 0 & 0 & 0 & 0 & 0 & -1 & 0 & 0\\
 0 & 0 & 0 & 0 & 0 & 0 & -1 & 0\\
 0 & 0 & 0 & 0 & 0 & 0 & 0 & 1 \end{array}\right),\quad
\underline e^2=\left( \begin{array}{llllllll}
 0 & -1 & 0 & 0 & 0 & 0 & 0 & 0\\
 1 & 0 & 0 & 0 & 0 & 0 & 0 & 0\\
 0 & 0 & 0 & -1 & 0 & 0 & 0 & 0\\
 0 & 0 & 1 & 0 & 0 & 0 & 0 & 0\\
 0 & 0 & 0 & 0 & 0 & 1 & 0 & 0\\
 0 & 0 & 0 & 0 & -1 & 0 & 0 & 0\\
 0 & 0 & 0 & 0 & 0 & 0 & 0 & 1\\
 0 & 0 & 0 & 0 & 0 & 0 & -1 & 0 \end{array}\right),
$$
$$
\underline e^3=\left( \begin{array}{llllllll}
 0 & i & 0 & 0 & 0 & 0 & 0 & 0\\
 i & 0 & 0 & 0 & 0 & 0 & 0 & 0\\
 0 & 0 & 0 & i & 0 & 0 & 0 & 0\\
 0 & 0 & i & 0 & 0 & 0 & 0 & 0\\
 0 & 0 & 0 & 0 & 0 & i & 0 & 0\\
 0 & 0 & 0 & 0 & i & 0 & 0 & 0\\
 0 & 0 & 0 & 0 & 0 & 0 & 0 & i\\
 0 & 0 & 0 & 0 & 0 & 0 & i & 0 \end{array}\right),\quad
\underline e^4=\left( \begin{array}{llllllll}
 0 & 0 & -1 & 0 & 0 & 0 & 0 & 0\\
 0 & 0 & 0 & 1 & 0 & 0 & 0 & 0\\
 1 & 0 & 0 & 0 & 0 & 0 & 0 & 0\\
 0 & -1 & 0 & 0 & 0 & 0 & 0 & 0\\
 0 & 0 & 0 & 0 & 0 & 0 & -1 & 0\\
 0 & 0 & 0 & 0 & 0 & 0 & 0 & 1\\
 0 & 0 & 0 & 0 & 1 & 0 & 0 & 0\\
 0 & 0 & 0 & 0 & 0 & -1 & 0 & 0 \end{array}\right),
$$
$$
\underline e^5=\left( \begin{array}{llllllll}
 0 & 0 & -i & 0 & 0 & 0 & 0 & 0\\
 0 & 0 & 0 & i & 0 & 0 & 0 & 0\\
 -i & 0 & 0 & 0 & 0 & 0 & 0 & 0\\
 0 & i & 0 & 0 & 0 & 0 & 0 & 0\\
 0 & 0 & 0 & 0 & 0 & 0 & -i & 0\\
 0 & 0 & 0 & 0 & 0 & 0 & 0 & i\\
 0 & 0 & 0 & 0 & -i & 0 & 0 & 0\\
 0 & 0 & 0 & 0 & 0 & i & 0 & 0 \end{array}\right).
$$

For
$(p,q)=(0,5)$ we take
$t=1/2(e+ie^1)1/2(e+ie^{23})$ and we have the same basis as for the signature $(3,2)$.
For the generators $e^2,e^3,e^4,e^5$ we have the same representations as for the case
$(1,4)$
and for the first generator we have
$$
\underline e^1=\left( \begin{array}{llllllll}
 -i & 0 & 0 & 0 & 0 & 0 & 0 & 0\\
 0 & i & 0 & 0 & 0 & 0 & 0 & 0\\
 0 & 0 & i & 0 & 0 & 0 & 0 & 0\\
 0 & 0 & 0 & -i & 0 & 0 & 0 & 0\\
 0 & 0 & 0 & 0 & -i & 0 & 0 & 0\\
 0 & 0 & 0 & 0 & 0 & i & 0 & 0\\
 0 & 0 & 0 & 0 & 0 & 0 & i & 0\\
 0 & 0 & 0 & 0 & 0 & 0 & 0 & -i \end{array}\right).
$$


\section{Unitary groups of Clifford algebras}
Consider the set of Clifford algebra elements
$$
\U\cl(p,q)=\{U\in\cl(p,q) : U^\dagger U=e\}.
$$
This set is closed with respect to the Clifford product and forms a group
(Lie group), which is called {\it the unitary group of Clifford algebra}.

Let $t\in\cl(p,q)$ be a Hermitian idempotent, $I(t)$ be the left ideal, and
$\tau_k$ be the orthonormal basis of $I(t)$. This basis gives us the matrix representation
of Clifford algebra elements (see Theorem 8)
$$
\gamma : \cl(p,q)\to\Mat(2^{[\frac{n+1}{2}]},\C)
$$
such that
$$
(\gamma(U))^\dagger=\gamma(U^\dagger),\quad \forall\,U\in\cl(p,q).
$$
In particular, we may take the standard basis and the standard
matrix representation from the previous section.

Let us take an element $U\in\U\cl(p,q)$ and the matrix $\gamma(U)\in
\Mat(2^{[\frac{n+1}{2}]},\C)$
\begin{equation}
U\tau_k=\gamma(U)_k^l\tau_l.
\label{Utau}
\end{equation}
The properties $U^\dagger U=e$ and $(\gamma(U))^\dagger=\gamma(U^\dagger)$ leads to the
property
$\gamma(U)^\dagger\gamma(U)={\bf 1}$, where ${\bf 1}$ is the identity matrix of dimension
$2^{[\frac{n+1}{2}]}$.
That means $\gamma(U)$ is a unitary matrix.

For even $n=p+q$ formula (\ref{Utau}) establishes the isomorphism
$$
\U\cl(p,q)\sim\U(2^{\frac{n}{2}}),
$$
where $\U(2^{\frac{n}{2}})$ is the group of unitary matrices of dimension
$2^{\frac{n}{2}}$.

For odd $n=p+q$ formula (\ref{Utau}) establishes the isomorphism
$$
\U\cl(p,q)\sim\U(2^{\frac{n-1}{2}})\oplus\U(2^{\frac{n-1}{2}}),
$$
where $\U(2^{\frac{n-1}{2}})\oplus\U(2^{\frac{n-1}{2}})$ is the set of block-diagonal
matrices $\diag(W,V)$ and $W,V\in\U(2^{\frac{n-1}{2}})$.

With the aid of an element $U\in\U\cl(p,q)$ we may define a new orthonormal basis of
$I(t)$
$$
\acute\tau_k=U\tau_k=\gamma(U)_k^l\tau_l,
$$
$$
(\acute\tau_k,\acute\tau^l)=\Tr((U\tau_k)^\dagger U\tau^l)=
\Tr(\tau_k^\dagger\tau^l)=\delta^l_k.
$$
The basis $\acute\tau_k=\acute\tau^k$ defines the new matrix representation
$\acute\gamma:\cl(p,q)\to\Mat(2^{\frac{n+1}{2}},\C)$
$$
V\acute\tau_k=\acute\gamma(V)^l_k\acute\tau_l.
$$
The representations $\gamma(V)$ and $\acute\gamma(V)$ are connected with each other by
the formula
$$
\acute\gamma(V)=\gamma(U)^{-1}\gamma(V)\gamma(U).
$$

If we replace the orthonormal basis $\tau_k$ of left ideal $I(t)$ by
the orthonormal basis $\check\tau_k=\tau_k U^{-1}$ of left ideal
$I(U t U^{-1})$, then we get the matrix representation
$V\to\check\gamma(V)$
$$
V\check\tau_k=V\tau_k U^{-1}=\check\gamma(V)_k^l\check\tau_l=
\check\gamma(V)^l_k\tau_l U^{-1}.
$$
Comparing this formula with formula (\ref{Utau}), we see that
$$
\check\gamma(V)=\gamma(V).
$$

Finally, if we replace the orthonormal basis $\tau_k$ of left ideal
$I(t)$ by the orthonormal basis $\hat\tau_k=U\tau_k U^{-1}$
of left ideal $I(U t U^{-1})$, then we get the matrix representation
$V\to\hat\gamma(V)$
$$
\hat\gamma(V)=\gamma(U^{-1})\gamma(V)\gamma(U).
$$

If the initial matrix representation
$\gamma : \cl(p,q)\to\Mat(2^{[\frac{n}{2}]},\C)$ is normal, then the matrix
representations
$\acute\gamma,\check\gamma,\hat\gamma$ are also normal.

\end{document}